\documentclass[preprint,nopreprintline,12pt]{elsarticle}

\usepackage{amssymb}
\usepackage{verbatim}
\usepackage{subfigure}
\usepackage{graphicx}
\usepackage{bm}
\usepackage{braket}
\usepackage{makecell}
\usepackage{amsthm}
\usepackage{amsmath}
\usepackage{amstext}
\usepackage{CJK}
\usepackage{diagbox}
\usepackage{cleveref}
\usepackage{array}
\usepackage{subfigure}
\usepackage{booktabs}
\usepackage{multirow}

\begin{document}

\begin{frontmatter}

\title{Quantum Recurrent Neural Networks for Sequential Learning}
\author{Yanan Li\fnref{fn1}}
\author{Zhimin Wang\fnref{fn1}\corref{cor1}}
\ead{wangzhimin@ouc.edu.cn}
\author{Rongbing Han}
\author{Shangshang Shi}
\author{Jiaxin Li}
\author{Ruimin Shang}
\author{Haiyong Zheng}
\author{Guoqiang Zhong}
\author{Yongjian Gu\corref{cor1}}
\ead{yjgu@ouc.edu.cn}
\fntext[fn1]{These authors contribute equally to this work.}
\cortext[cor1]{Co-corresponding author}
\affiliation{organization={Faculty of Information Science and Engineering, Ocean University of China},
            city={Qingdao},
            postcode={266100},
            country={China}}

\begin{abstract}
Quantum neural network (QNN) is one of the promising directions where the near-term noisy intermediate-scale quantum (NISQ) devices could find advantageous applications against classical resources. Recurrent neural networks are the most fundamental networks for sequential learning, but up to now there is still a lack of canonical model of quantum recurrent neural network (QRNN), which certainly restricts the research in the field of quantum deep learning. In the present work, we propose a new kind of QRNN which would be a good candidate as the canonical QRNN model, where, the quantum recurrent blocks (QRBs) are constructed in the hardware-efficient way, and the QRNN is built by stacking the QRBs in a staggered way that can greatly reduce the algorithm’s requirement with regard to the coherent time of quantum devices. That is, our QRNN is much more accessible on NISQ devices. Furthermore, the performance of the present QRNN model is verified concretely using three different kinds of classical sequential data, i.e., meteorological indicators, stock price, and text categorization. The numerical experiments show that our QRNN achieves much better performance in prediction (classification) accuracy against the classical RNN and state-of-the-art QNN models for sequential learning, and can predict the changing details of temporal sequence data. The practical circuit structure and superior performance indicate that the present QRNN is a promising learning model to find quantum advantageous applications in the near term.
\end{abstract}


\end{frontmatter}


\section{Introduction}
In recent years, deep neural networks (DNNs) \cite{lecun2015a} have enabled revolutionary applications in several domains of artificial intelligence \cite{jordan2015}, such as computer vision \cite{voulodimos2018} and natural language processing \cite{goldberg2016}, etc. Parallelly, remarkable breakthroughs have been seen in quantum computing \cite{bharti2022b,madsen2022,zhong2020a,arute2019a}. With the demonstrations of quantum supremacy, we are entering the NISQ era of quantum computing, where NISQ refers to the noisy intermediate-scale quantum devices \cite{preskill2018}. There is a growing consensus that NISQ devices may find useful applications in the near term. One of the most promising directions is the quantum neural network (QNN) \cite{mitarai2018b,schuld2020a,farhi2018b}. QNN takes the parameterized quantum circuit (PQC) \cite{benedetti2019a} as a learning model, which is a quantum analogue of the classical neural network.   

The great success of classical DNNs is mainly attributed to its flexible architecture. That is, the multilayer architecture is versatile to discover intricate structures in high-dimensional data. Specifically, the convolutional neural networks (CNNs) \cite{gu2018} can effectively capture spatial correlation within, say, image data; while the recurrent neural networks (RNNs) perform well when learning sequential data \cite{graves2014}, e.g., natural language processing. 

Inspired by DNNs, naturally quantum deep neural networks (QDNNs) should have similar architectures to process corresponding types of data. Indeed, for the quantum convolutional neural networks (QCNNs), there have been a fair amount of research covering QCNNs' structure, learnability, and applications \cite{cong2019,pesah2021,hur2022c,herrmann2022}. In contrast, the studies about quantum recurrent neural networks (QRNNs) are rather sparse. Bausch developed a high-degree nonlinear quantum neuron \cite{bausch2020b} based on the work of Cao et al.~\cite{cao2017a}, and used this neuron to build the recurrent networks. Such models possess good non-linearity but need to implement amplitude amplification operations, resulting in high complexity of quantum circuits. Takaki et al. \cite{takaki2021b} proposed a kind of QRNNs by employing a PQC with a recurrent structure. Such models use simple quantum circuits that are easy to implement on the NISQ devices, but the performance on the non-trivial sequential data has yet to be verified. In addition, Sipio and Chen et al. \cite{sipio2022,chen2020a} developed hybrid quantum-classical models of QRNNs, where the classical linear layers in RNNs are replaced with PQCs. Such hybrid models just take the quantum circuits as acceleration sub-modules plugging into the classical networks, and face with the dilemma of the interface between quantum and classical systems.   

Until now there is still a lack of canonical model of QRNN, which certainly restricts the research of QRNNs. Inspired by the QCNN model proposed by Cong et al. \cite{cong2019}, we consider that one canonical QRNN model should possesses the following features: (1) flexible to be implemented on various NISQ platforms; (2) fully quantum evolved but not the quantum-classical hybrid networks to ease the interface problem; (3) efficient for sequential learning of classical data.

In order to address the above issues, in the present work, we develop a new kind of QRNN that can fulfill the requirements of the canonical QRNN. Specifically, 
\begin{enumerate}
\item We propose to construct the quantum recurrent block in a more hardware-efficient way, but not based on Hamiltonian dynamics as done by Takaki et al. \cite{takaki2021b}. More importantly, we propose a staggered architecture of QRNN by stacking the recurrent blocks in a staggered way. The staggered QRNN can greatly reduce the algorithm's requirement with regard to the coherent time of quantum devices. This property is of great significance, because increasing the coherent time of quantum hardware is extremely hard from the technology development point of view.
\item The present QRNN is a fully quantum learning model, where the outcome of quantum transformation is taken as the prediction of the data with minor post-processing. Our QRNN is a standard NISQ algorithm \cite{bharti2022b}, which can take full advantage of the near term quantum computers.
\item The performance of the present QRNN on classical sequential data is verified concretely. Three different kinds of sequential data including meteorological indicators, stock price, and natural language are applied to test the models. Our QRNN model shows better performance in prediction (classification) accuracy against the classical RNN and state-of-the-art QNN models for sequential learning, and can predict the changing details of the sequence data.  
\end{enumerate}
The simple structure as well as the good performance imply that our QRNN would be a promising candidate to find useful applications in the near term.

The rest of the paper is organized as follows. In Section~\ref{sec:2}, we describe the structure of the quantum recurrent block, which is the basic cell to construct the QRNNs. Section~\ref{sec:3} shows the details of the QRNNs, including the architectures of QRNN and the method of optimizing parameters. In Section~\ref{sec:4}, we present the performance of the QRNNs on three kinds of classical sequential data, i.e., the data of meteorological indicators, stock price, and natural language. Finally, conclusions and outlook of the present work are discussed in Section~\ref{sec:5}.

\section{Quantum recurrent block}\label{sec:2}
In general, RNNs possess a multilayer architecture and each layer is the basic recurrent block. Depending on the specific design of the recurrent block, there are a number of RNN variants, such as long short-term memory (LSTM) \cite{hochreiter1997} and gated recurrent unit (GRU) \cite{cho2014}. The main idea of the recurrent block is that the prediction at a given moment is determined by both the new input data of the current moment and the information about the history of all the past elements of the sequence.

\begin{figure}[ht]
	\centering
	\includegraphics[width=\linewidth]{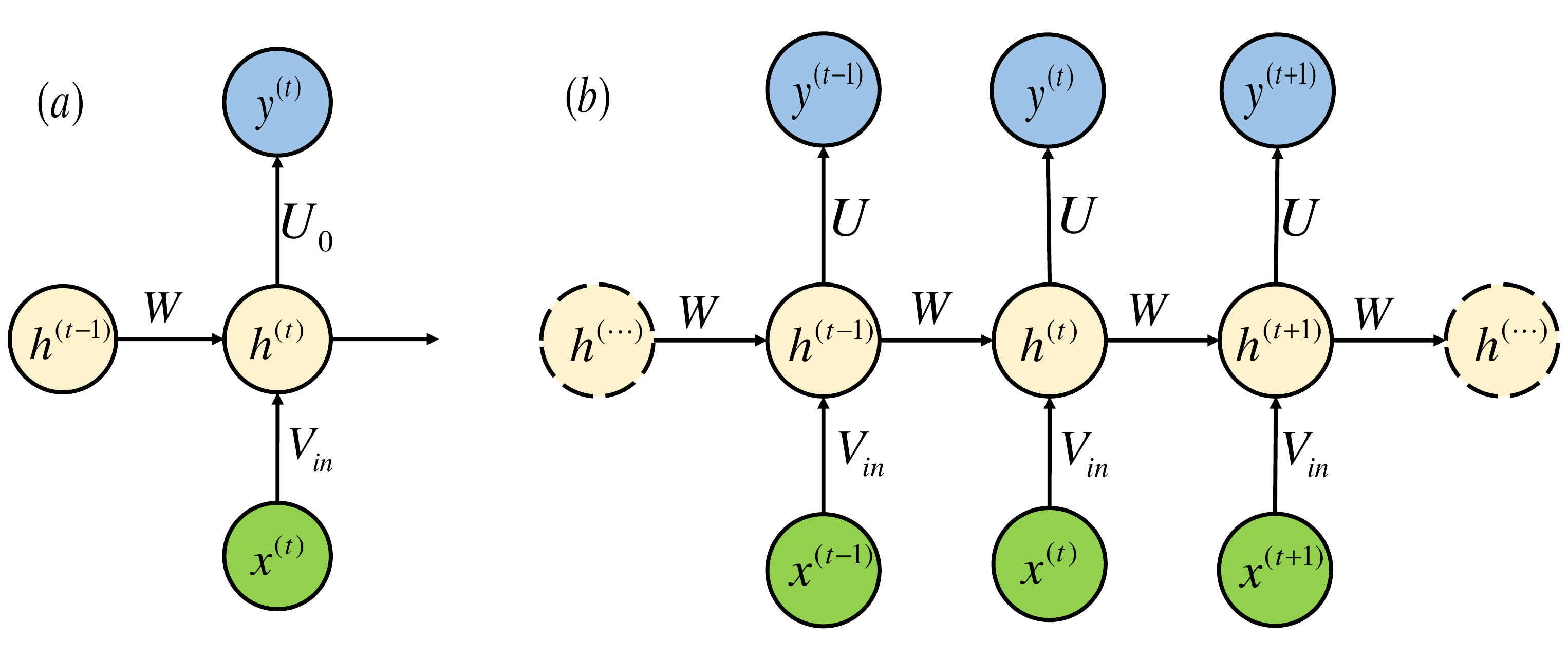}
	\caption{(a) Structure of the basic classical recurrent block. (b) One basic architecture of RNN by stacking the basic recurrent blocks.}
	\label{fig:1}
\end{figure}

In the most basic recurrent block, the output at the time step $t$ can be expressed as 
\begin{eqnarray}
	\begin{aligned}
		&\vec{y}^{(t)} = f_y (U_o \vec{h}^{(t)}), \\
		&\vec{h}^{(t)} = f_h (V_{in} \vec{x}^{(t)} + W\vec{h}^{(t-1)}),
	\end{aligned}
\end{eqnarray}
where $\vec{x}^{(t)}$ is the input of time step $t$, $\vec{h}^{(t-1)}$ is the output of last step implicitly containing information about all the previous elements of the sequence, $f_y$ and $f_h$ are the activation functions. The training process of RNNs is to optimize the parameters $V_{in}$, $W$ and $V_o$ by minimizing a loss function. The basic classical recurrent block and one RNN by stacking the recurrent blocks are schematically shown in Fig.~\ref{fig:1}.

\begin{figure}[ht]
	\centering
	\includegraphics[width=\linewidth]{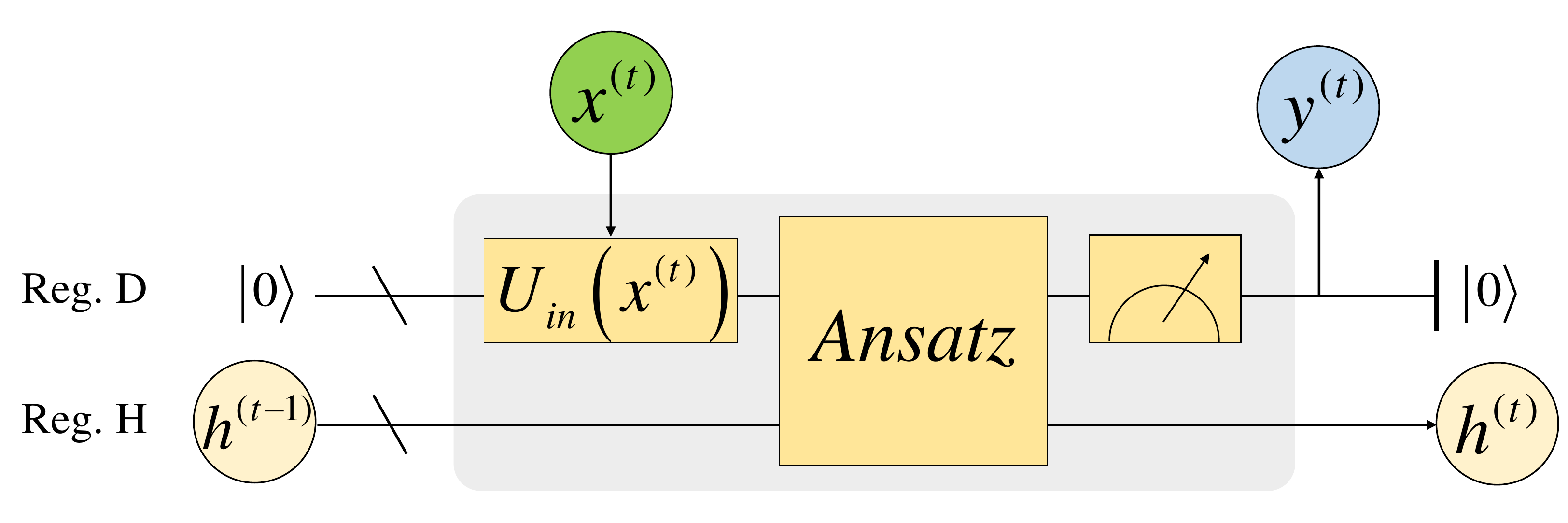}
	\caption{Structure of the quantum recurrent block inspired by the basic classical recurrent block shown in Fig.~\ref{fig:1}.}
	\label{fig:2}
\end{figure}

Inspired by the structure of the above classical recurrent block, the quantum recurrent block (QRB) is designed as schematically shown in Fig.~\ref{fig:2}. The qubits of QRB are divided into two groups, i.e., two quantum registers denoting as Reg. D and Reg. H. Reg. D is used to embed the sequential data, one element at each time step, and Reg. H is used to store information about the history of all previous elements. In general, the QRB consists of three parts, which are data encoding $U_{in}(x^{(t)})$, ansatz circuit, and partial quantum measurement. Below we go into the details of implementing the three parts.

\subsection{Data encoding}
Data encoding is a process of loading classical data into quantum computer, which is to represent the classical data as quantum states. It is worth noting that in quantum machine learning, data encoding plays a crucial role, far beyond the thing of preparing the input. On the one hand, encoding the classical data quantumly is by no means a trivial thing, which would turn to be the bottleneck for the runtime of the whole algorithm. On the other hand, data encoding can be interpreted as a feature map, dubbed as ``quantum feature map'', which is to map the input to Hilbert space of the quantum system \cite{schuld2021b}. A well-chosen quantum feature map can make the data linearly separable in the feature space, and thereby efficiently solve the learning problem.

Data encoding is equivalent to performing a unitary transformation $U_{in}(x)$ on the initial state, i.e., $\ket{f(x)}=U_{in}\ket{0}^{\otimes n}$ with $n$ being the number of qubits. There exist numerous structures of circuit to embed the classical data into a quantum state. Among them, the most famous one would be the amplitude encoding, which can embed exponentially many classical data \cite{schuld2021b}. Specifically, given a normalized classical vector $x=(x_1,\dots,x_N)^T$ of dimension $N = 2^n$, amplitude encoding can represent this vector as amplitudes of an $n$-qubit quantum state, i.e., $U_{in}(x)\ket{0}^{\otimes n} = \sum_{i=0}^{2^n-1}x_i\ket{i}$. Similarly, a data matrix $A\in\mathbb{C}^{2^n\times2^n}$ with entries $a_{ij}$ satisfying $\sum_{ij}|a_{ij}|^2=1$ can be encoded as $U_{in}(A)\ket{0}^{\otimes m}\ket{0}^{\otimes n} = \sum_{i=0}^{2^m-1}\sum_{j=1}^{2^n-1}a_{ij}\ket{i}\ket{j}$, where $\ket{i}$ and $\ket{j}$ are respectively the $i$th and $j$th computational basis state. However, amplitude encoding is much less common in QNNs, because the quantum circuit cost for amplitude encoding usually grows as $O(poly(N))$ that is exponential of the number of qubits. Therefore, although having exponentially large data-encoding space, amplitude encoding cannot be implemented efficiently on NISQ devices.

In QNNs, the most commonly used encoding techniques are the angle encoding \cite{schuld2021b} and the associated circuit encoding \cite{havlicek2019}. In general, angle encoding and circuit encoding embed the classical data as the rotation angles of the single-qubit or controlled two-qubit rotation gates. Specifically, given one classical data point $x_i$, the angle encoding first rescales the data to $\tilde{x}_i$ lying between $0$ and $\pi$, and then embeds it into a single qubit as, say $U_{in}(\tilde{x}_i)\ket{0}=\cos(\frac{\tilde{x}_i}{2})\ket{0}+\sin(\frac{\tilde{x}_i}{2})\ket{1}$ with $U_{in}(\tilde{x}_i)$ being the $R_y$ gate $R_y(\theta)=[\cos\frac{\theta}{2},-\sin\frac{\theta}{2},\sin\frac{\theta}{2},\cos\frac{\theta}{2}]$ \cite{nielsen2010}. For $N$ data points $x=(x_1,\dots, x_N)^T$, the angle encoding embeds them by $N$ qubits,
\begin{eqnarray}
	R_y^{\otimes N}(\tilde{x})\ket{0}^{\otimes N} = \bigotimes_{i=1}^N(\cos(\frac{\tilde{x}_i}{2})\ket{0} + \sin(\frac{\tilde{x}_i}{2})\ket{1}).
\end{eqnarray} 
Formally, angle encoding is actually a kind of time-evolution encoding. Time-evolution encoding prescribes to associate a scalar value $x\in R$ with the time $t$ in the unitary evolution by a Hamiltonian $\hat{H}$, i.e., $U(x)=exp(-i\hat{H}t)$. In the angle encoding, $\hat{H}$ is just the Pauli operation, and the corresponding unitaries $U(\theta)=exp(-i\hat{\sigma}\theta)$ can be implemented efficiently on NISQ devices. Therefore, the computational cost of angle encoding is minor, while the required number of qubit is $O(N)$ for $N$ data points.	

Circuit encoding takes the angle encoding as the basis, and embeds the data into a more complex circuit than that of angle encoding. For example, more than one data point, say ${x_1,x_2,x_3}$, can be encoded into one qubit as follows, 
\begin{eqnarray}
	R_z(\tilde{x}_3)R_x(\tilde{x}_2)R_z(\tilde{x}_1)H\ket{0}.
\end{eqnarray}
Such strategy of dense encoding can address, to some extent, the problem of angle encoding which needs $N$ qubits to embed $N$ data points. Based on this strategy, up to $67-$dimensional data has been handled successfully using quantum kernel method on the current quantum processors \cite{peters2021}. More importantly, circuit encoding can be used to construct a complex feature map which is hard to simulate by the classical computers. For example, in Ref. \cite{havlicek2019}, one encoder is proposed as follows,
\begin{eqnarray}
	U_{\phi(X)}=exp(i\sum_{j,k}^{n}\phi_{j,k}(x)Z_jZ_k)H^{\otimes n},
\end{eqnarray}
where $n$ is the number of qubit, $Z_j (Z_k)$ is the Pauli-Z operator for the $j$th ($k$th) qubit, $\phi_{j,k}$ are real functions, and $H$ is the Hadamard gate. Even two layers of such an encoder circuit would make it computationally hard for the classical resources \cite{havlicek2019}. However, it is worth mentioning that whether such a complex feature map leads to an advantage for discovering structure in data is still an open question. 

\begin{figure}[ht]
	\centering
	\includegraphics[scale=0.3]{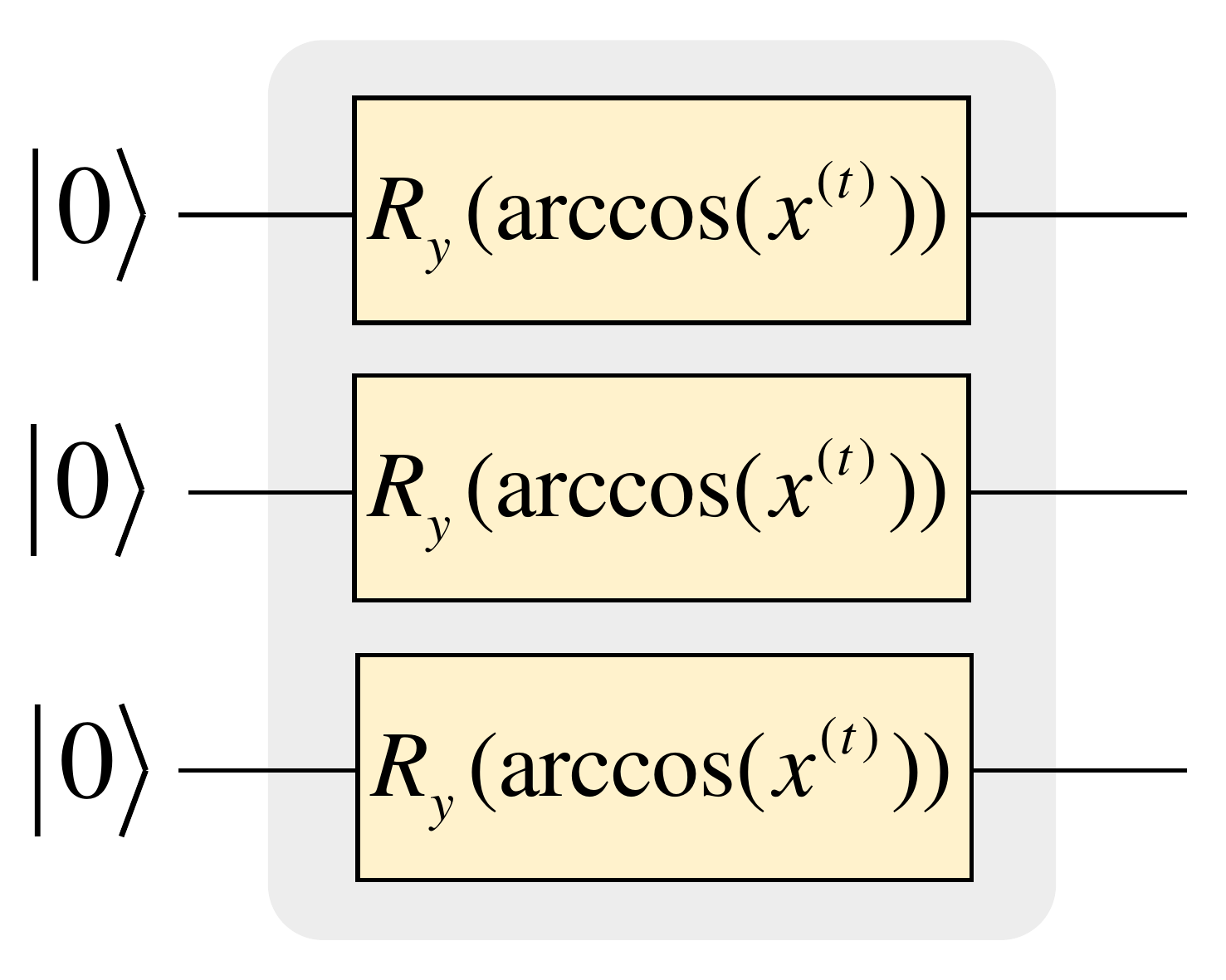}
	\caption{The circuit for encoding the sequential data at the time step $t$.}
	\label{fig:3}
\end{figure}

In the present work, we use the angle encoding to load the sequential data. This choice is mainly due to the fact that the scope of the present work is to verify whether the QRNNs we developed are efficient for learning classical sequential data. So we tend to employ the data encoding as simple as possible and especially the commonly used one on NISQ devices. Specifically, the circuit for angle encoding used in the present work is shown in Fig.~\ref{fig:3}. Note that each element of the sequential data is embedded in a replicative fashion. It has been shown that input redundancy can provide an advantage in classification accuracy \cite{gilvidal2020,peruzzo2014}.

Generally, more complex data encoding techniques, such as the circuit encoding, should be helpful to increase the performance of our QRNNs. We leave this exploration for future work. 

\subsection{Ansatz}
The ansatz in Fig.~\ref{fig:2} is to process the feature produced by the encoder circuit from the raw data, and output the new feature for the following task of classification or regression. Ansatz is, in fact, a parameterized quantum circuit, which has adjustable quantum gates. The adjustable parameters are optimized based on the data to learn, as they are determined in the algorithm of neural networks.

Ansatz, namely the PQC, is used to approximate the target function, which can map the feature data into different value domains representing different classifications. Similar to the universal approximation theorem in classical neural networks \cite{hornik1989}, there always exists a quantum circuit that can approximate the target function within an arbitrary small error \cite{cai2022a,schuld2021f}. Moreover, it is possible to construct such a circuit with polynomial cost of quantum gates \cite{benedetti2019a,peruzzo2014}. Therefore, the ansatz aims to use polynomial number of quantum gates (i.e., polynomial number of parameters) to implement a function that can approximate the task at hand.

In practice, ansatz follows a fixed structure of quantum gates. There are two strategies to design the circuits, which are problem-inspired and hardware-efficient methods. Problem-inspired method is to leverage the Hamiltonian of the problem to construct the circuit, and optimizing the parameters of the circuit always means to search the ground state of the Hamiltonian. The ansatz applied in the variational quantum algorithms (VQAs) for solving eigenvalues and eigenstates \cite{cerezo2021a,peruzzo2014}, and for approximation and optimization \cite{farhi2014} usually adopts the problem-inspired method. Whereas, the ansatz in QNNs is typically hardware-efficient circuit, which consists of layers of native entangling gates (i.e., two-qubit gates) and single-qubit gates \cite{kandala2017}. Essentially, such ansatz can be customized concerning the gate set and connectivity of the specific quantum devices, and the ansatz can be implemented directly on the device without the need of compilation.

Here, the ansatz circuit used in Fig.~\ref{fig:2} is constructed in the hardware-efficient way. Specifically, the circuit is composed of single-qubit rotation gates and two-qubit gates (i.e., controlled rotation gates). As it is known, an arbitrary single-qubit gate can be expressed as a combination of rotation gates about the $\hat{x},\hat{y}$ and $\hat{z}$ axes \cite{nielsen2010}. We adopt the $X-Z$ decomposition to represent the single-qubit gates in the circuit,
\begin{eqnarray}\label{eq:5}
	U_{1q} = R_x(\alpha)R_z(\beta)R_x(\gamma),
\end{eqnarray}     
where $\alpha$, $\beta$, and $\gamma$ are the adjustable parameters to be optimized in the learning process.

\begin{figure}[ht]
	\centering
	\includegraphics[scale=0.15]{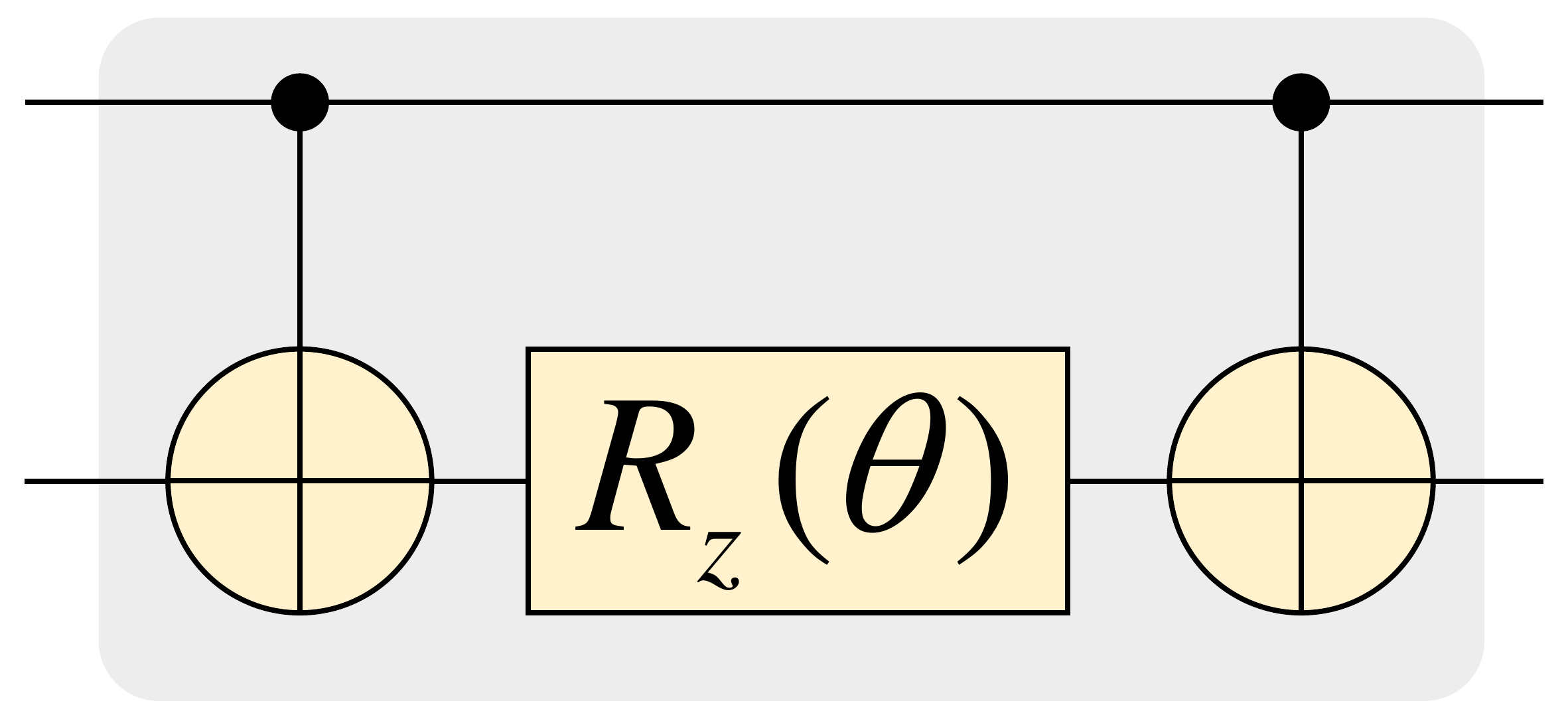}
	\caption{Implementation of the two-qubit gate $R_{zz}(\theta)$ using the CNOT and Pauli-Z gate.}
	\label{fig:4}
\end{figure}

For the two-qubit gates, they are applied to produce entanglement between qubits in the circuit. The two-qubit gates would be fixed without adjustable parameters, such as the CNOT and controlled Pauli-Z gate; or they would be adjustable, basically the controlled $R_x(\theta)$ and $R_z(\theta)$ gates. In order to increase the expressibility and entangling capability of the ansatz, we use the $R_{zz}(\theta)$ gate as the two-qubit gates in the circuit,
\begin{eqnarray}\label{eq:6}
	U_{2q} = R_{zz}(\theta) = exp(i\theta Z_j Z_k),
\end{eqnarray}
where $Z_j$ and $Z_k$ are the Pauli-Z operators on the $j$th and $k$th qubit, respectively. The  $R_{zz}(\theta)$ gate can be implemented using the CNOT and Pauli-Z gates as shown in Fig.~\ref{fig:4}.

\begin{figure}[ht]
	\centering
	\includegraphics[width=\linewidth]{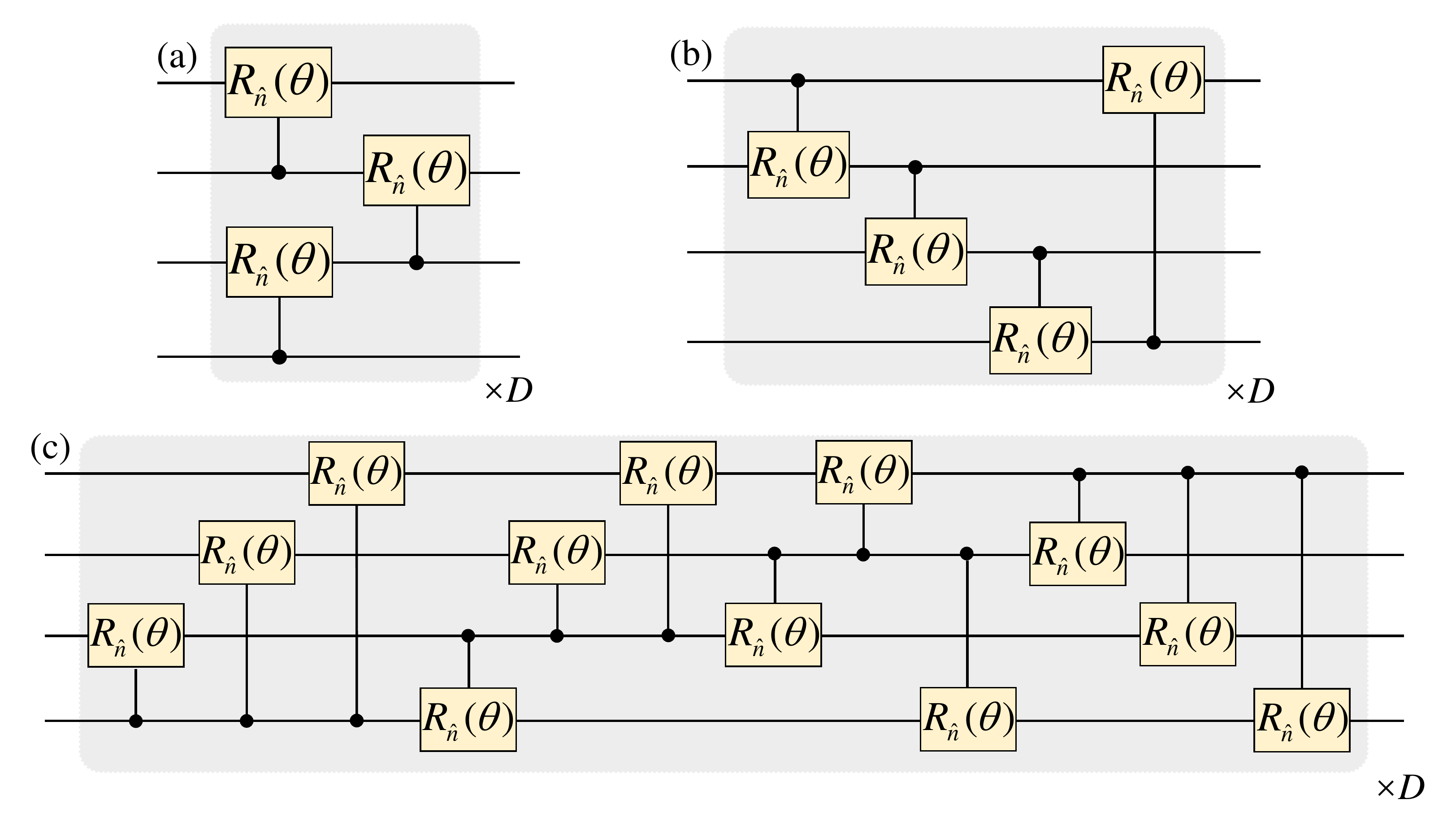}
	\caption{The circuits for the three configurations of two-qubit gates, (a) nearest-neighbor (NN), (b) circuit-block (CB), and (c) all-to-all (AA).}
	\label{fig:5}
\end{figure}

Next, the two-qubit gates need to be arranged in a regular way to boost the expressibility and entangling capability of the whole circuit. There are mainly three configurations of two-qubit gates, which are nearest-neighbor (NN), circuit-block (CB), and all-to-all (AA) structures as shown in Fig.~\ref{fig:5}. On the one hand, these configurations are proposed to harness various quantum hardware platforms with different qubit topologies. Specifically, the NN circuit is the most natural configuration for quantum devices with a linear array of qubits, while AA requires a fully connected architecture of qubits. On the other hand, as discussed in Ref. \cite{sim2019a}, the three configurations are distinguished from each other in the properties of expressibility, entangling capability, and circuit cost. When having the same number of two-qubit gates (e.g., $D = 4$ for NN, $D = 3$ for CB, and $D = 1$ for AA in Fig.~\ref{fig:5}), NN circuit has the worst expressibility and entangling capability, but the lowest circuit depth; while AA has the best expressibility and entangling capability, but the highest requirement of circuit depth and connectivity. The CB circuit can provide a good balance. CB has a much cheaper circuit, while its expressibility and entangling capability are slightly less and equal to AA. Therefore, we apply the CB configuration in the ansatz circuit.

\begin{figure}[ht]
	\centering
	\includegraphics[width=\linewidth]{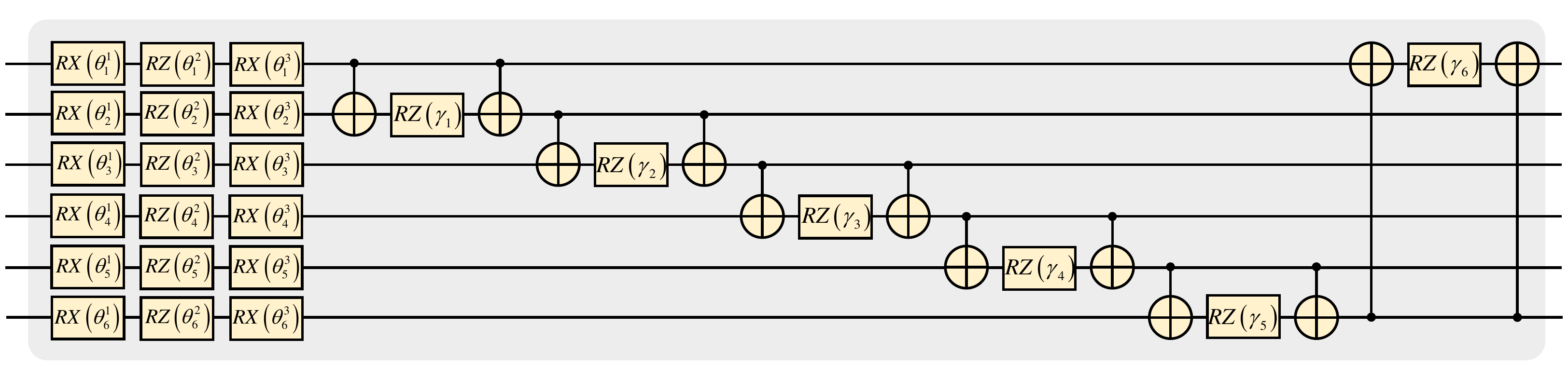}
	\caption{The circuit of the ansatz used in Fig.~\ref{fig:2}.}
	\label{fig:6}
\end{figure}

Finally, we put all the things together, including the single-qubit gates in Equation~\ref{eq:5}, the two-qubit gates in Fig.~\ref{fig:4} and the CB configuration of two-qubit gates in Fig.~\ref{fig:5}, then the circuit of the ansatz in Fig.~\ref{fig:2} is obtained. Fig.~\ref{fig:6} shows one ansatz circuit with $6$ qubits. 

\subsection{Partial quantum measurement}
The final step of the recurrent block is to output a prediction $y_t$ of the current stage, and maintain an intermediate state $h_t$ that contains information about the history of the sequential data. This is achieved by implementing the partial quantum measurement, namely measuring a portion of qubits as shown in Fig.~\ref{fig:2}. 

Quantum measurement corresponds to a physical observable $M$, which can be decomposed as $M=\sum_i \lambda_i P_i$ with $\lambda_i$ being the $i$th eigenvalue and $P_i$ the projector on the corresponding eigenspace. According to the Born rule, the outcome of the measurement corresponds to one of the eigenvalues $\lambda_i$; that is, the quantum state of the qubits $\ket{\varphi}$ randomly collapse to the corresponding eigenstates $\ket{\lambda_i}$ with a probability $p(\lambda_i)=\bra{\varphi}P_i\ket{\varphi}$. Then, the expectation value of the measurement outcome can be formalized as
\begin{eqnarray}
	\left\langle M \right\rangle = \sum_i \lambda_i p(\lambda_i) = \sum_i \lambda_i\bra{\varphi}P_i\ket{\varphi}.
\end{eqnarray}

The most straightforward and commonly used measurement in quantum algorithms is the computational basis measurement, namely the Pauli-Z measurement with the observable $\sigma_Z$,
\begin{eqnarray}
	\sigma_Z = (+1)\ket{0}\bra{0}+(-1)\ket{1}\bra{1}.
\end{eqnarray}
When acting on multiple qubits, Pauli-Z measurement measures whether the individual qubits are in state $\ket{0}$ or $\ket{1}$, from which we can read off the eigenvalues $+1$ or $-1$, respectively. 

The expectation value $\langle \sigma_Z \rangle$ of a single-qubit is a value in the range $[-1, 1]$. Note that every time we measure a quantum state $\ket{\varphi}$, it will collapse to $\ket{0}$ or $\ket{1}$. Hence, in practice, the expectation is estimated by repeating the operations of creating the state $\ket{\varphi}$ and measuring for $S$ time, where $S$ is also known as the number of shots. The average of the $S$ results is taken as an estimate of the expectation. Therefore, high-precision estimates of the expectation require a larger number of shot, namely rerunning the algorithm many times. It can be proved that the scaling of $S$ is $O((1/\varepsilon^2))$, where $\varepsilon$ is the error of the estimation \cite{schuld2021e}. 

In the partial quantum measurement as shwon in Fig.~\ref{fig:2}, only the quantum Reg. D is measured. Specifically, we first implement the Pauli-Z measurement only on the first qubit of Reg. D. The probability of the first qubit collapsing to state $\ket{1}$ is estimated, and after minor post-processing it is taken as the prediction $y_t$. Then, all the qubits of Reg. D are measured and reinitialized to the state $\ket{0}$ to be ready to embed the next element of the sequential data. Here, we would like to mention that the probability of the first qubit collapsing to $\ket{1}$ can be expressed using the partial trace technique. Specifically, passing through the circuits of data encoding $U_{in}$ and ansatz $U_a$ , the initial state evolves into $U_aU_{in}\ket{0}$, and the corresponding dense matrix is $\rho = U_aU_{in}\ket{0}\bra{0}U_{in}^\dagger U_a^\dagger$. Then the reduced density operator of the first qubit of Reg. D is $\rho^{(1)}=tr_{\bar{1}}(\rho)$, where $tr_{\bar{1}}$ represents the partial trace over the left qubits of Reg. D except the first qubit. Considering that the measurement operator acting on the first qubit is the projector $\ket{1}\bra{1}$, then the probability is 
\begin{eqnarray}
	p(\ket{1})=tr\left(\ket{1}\bra{1}\rho^{(1)}\right).
\end{eqnarray}
We just use the partial trace technique to implement the partial quantum measurement in our program for simulating the present QRNN.

There are two reasons why only the first qubit, rather than the global qubits of Reg. D is measured. First, less qubits to measure can greatly reduce the measurement error when implementing the algorithm on quantum devices. Second, global measurement would exhibit a barren plateau (BP), that is, the cost function gradients would vanish exponentially in the number of qubits \cite{cerezo2021b}. The partial measurement used here can reduce the BP, and thus improve the trainability of the network. 

\section{Quantum Recurrent Neural Networks}\label{sec:3}
After having the quantum recurrent block, we can construct the quantum recurrent neural networks immediately by stacking the blocks with certain rules. Below we first present two architectures of QRNNs, then discuss the way of optimizing the parameters of QRNNs.

\subsection{Two QRNN architectures}
\begin{figure}[ht]
	\centering
	\includegraphics[width=\linewidth]{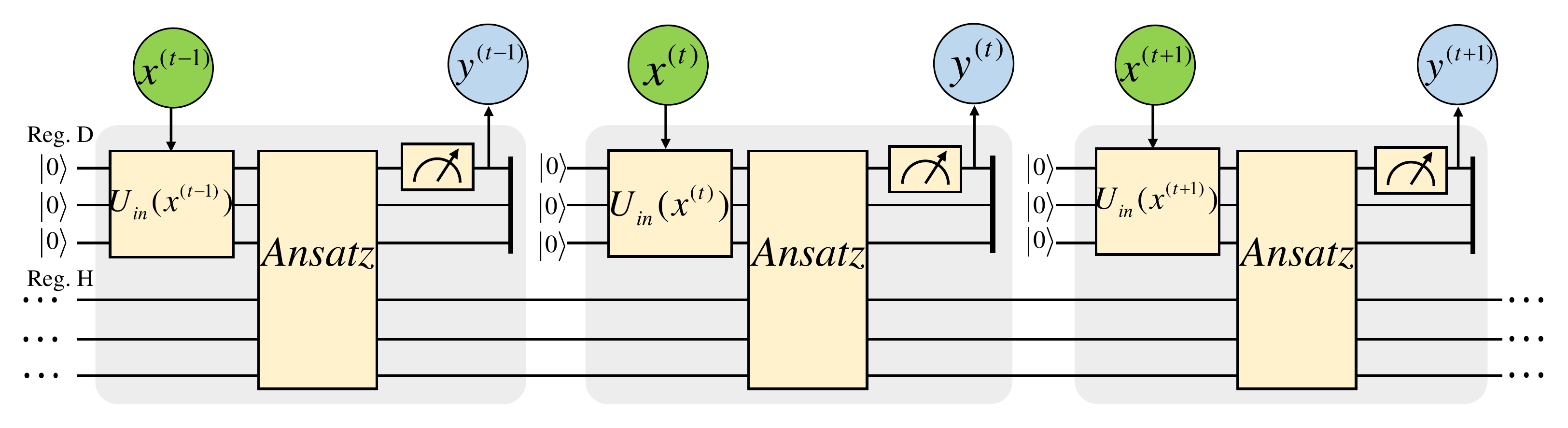}
	\caption{The straightforward way of building the QRNNs using the QRB. This architecture is called as pQRNN (i.e., plain QRNN).}
	\label{fig:7}
\end{figure}

The most straightforward way of building the QRNNs is to arrange the QRB in sequence as shown in Fig.~\ref{fig:7}. Hereafter, this architecture is called as pQRNN, i.e., plain QRNN. As discussed in Section~\ref{sec:2}, each QRB goes as follows: first Reg. D embeds one element of the sequential data into the circuit; then Reg. D and Reg. H are entangled through the ansatz; further the first qubit of Reg. D is measured to get an intermediate prediction; finally Reg. D reinitializes to the state $\ket{0}$ and Reg. H feeds its state directly into the next QRB to transmit the history information of the sequential data. 

In pQRNN, qubits assigned to Reg. D and Reg. H are fixed. That is, when implementing pQRNN, qubits of Reg. H need to work all the time. Hence, pQRNN requires the quantum devices having a long coherent time. However, coherent time is one of the central indicators of NISQ devices, which is extremely hard to increase greatly. 

\begin{figure}[ht]
	\centering
	\includegraphics[width=\linewidth]{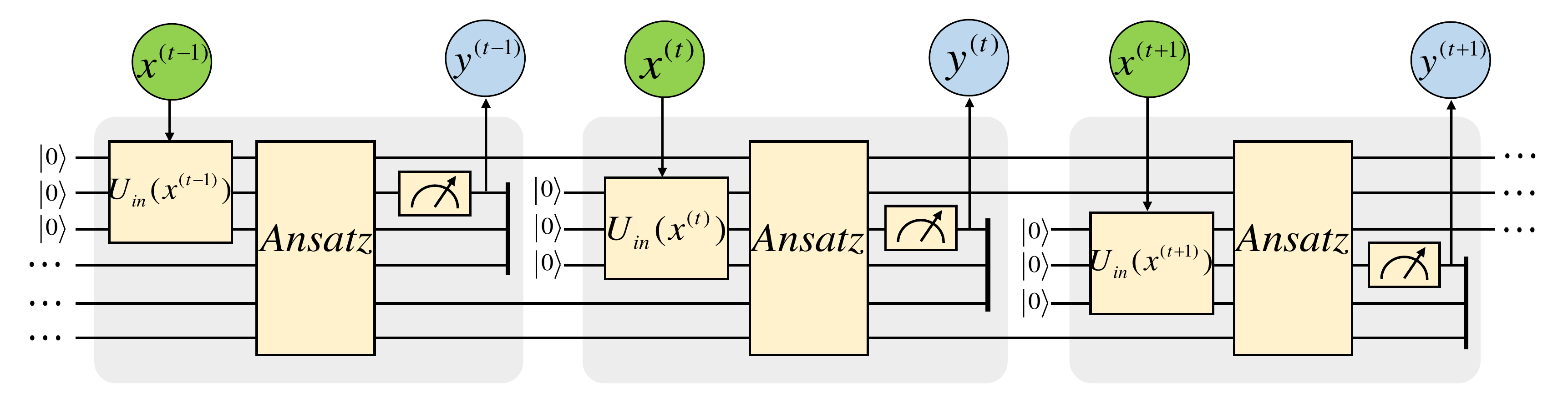}
	\caption{The second architecture of QRNN, which is built by arranging the QRB in a staggered way. This model is called as sQRNN (i.e., staggered QRNN).}
	\label{fig:8}
\end{figure}

In order to address the issues of pQRNN, we propose another architecture of QRNN. In this model, the QRBs are arranged in a staggered way as shown in Fig.~\ref{fig:8}, and it is called as sQRNN, i.e., staggered QRNN. In sQRNN, qubits are in turn assigned to Reg. H, so each qubit has a chance to reinitialize to state $\ket{0}$ after several time steps. Using the strategy of shift work, sQRNN greatly reduces the requirement of coherent time of quantum hardware, therefore being more accessible on near-term quantum devices.

\subsection{Parameters learning}
With the two architectures of QRNNs, now we discuss the method of learning optimal parameters of QRNNs. First, predictions (i.e., outcome of quantum measurement) of different discrete time steps are rescaled to associate with the real values or labels. The rescaled predictions are
\begin{eqnarray}
	\tilde{y}_t = y_t \cdot (x_{max} - x_{min}) + x_{min},
\end{eqnarray}
where $x_{min}$ and $x_{max}$ are the minimum and maximum of input states. 

Just like classical neural networks, the errors between the predictions and the real values are quantified by the loss functions. Learning parameters is done by minimizing the errors. There are various loss functions to use, which create different landscapes with different properties of plateaus, saddle points, and global minima, etc. The widely used methods include the mean squared error (i.e., $L_2$ loss) and the cross-entropy loss. In the present work, we use the most straightforward $L_2$ loss,
\begin{eqnarray}
	L_2(\vec{\theta}) = \frac{1}{N}\left(\tilde{y}_t(x,\vec{\theta}) - y_{true}\right)^2,
\end{eqnarray}
where $\vec{\theta}$ are the parameters to learn (i.e., the rotation angles in ansatz circuit) and $N$ is the number of data samples. 

Just like classical neural networks, the parameters can be optimized based on the gradient of the loss function. That is, the parameters are updated towards the direction of steepest descent of the loss function. Gradient-based approaches play a crucial role in deep learning, because there exists the backpropagation (BP) algorithm, which can implement the calculation of the derivatives of the loss function with respect to the parameters through a computationally efficient organization of the chain rule. However, in quantum computing, there exists no similar BP algorithm to evaluate the derivatives. This is because the BP algorithm relies on storing the intermediate state of the network during computation, which is forbidden by the quantum no-cloning theorem. 

In quantum deep learning, there are two typical methods to compute the derivatives of the parameters, namely the difference method and the analytical method. In the difference method, the partial derivatives are approximated by the finite difference scheme,
\begin{eqnarray}\label{eq:12}
	\frac{\partial L_2(\vec{\theta})}{\partial \theta_j} = \frac{L_2(\vec{\theta} + \Delta \cdot \vec{e_j}) - L_2(\vec{\theta} - \Delta \cdot \vec{e_j})}{2 \Delta},
\end{eqnarray}
where $\Delta$ is a tiny hyper-parameter, and $\vec{e_j}$ is an unit vector with the $j$th element being $1$ and the rest being $0$. Note that in order to estimate the derivative of each parameter, this method requires to evaluate the loss function twice, that is, implementing the quantum circuit twice. 

In the analytical method, analytical gradients can be obtained based on the feature of the quantum gate used in the PQC. Suppose the trainable gates in PQC are of the form $U(\theta_j)=e^{-i\theta_jP_j}$, where $P_j$ is a tensor product of Pauli matrices (in fact, almost all of the PQCs in literature apply this form of quantum gates). Then the derivative of the measurement expectation value (i.e., Equation~\ref{eq:6}) respect to the parameter $\theta_j$ can be formalized as \cite{benedetti2019a} 
\begin{eqnarray}\label{eq:13}
	\frac{\partial\langle M \rangle}{\partial \theta_j} = \frac{\langle M \rangle_{\vec{\theta} + \frac{\pi}{2} \cdot \vec{e_j}} - \langle M \rangle_{\vec{\theta} - \frac{\pi}{2} \cdot \vec{e_j} }} {2},
\end{eqnarray}
where the subscript $\vec{\theta} \pm \frac{\pi}{2} \cdot \vec{e_j}$ indicate that the parameter $\theta_j$ add (minus) $\pi/2$. That is, the derivative is estimated by executing the two circuits with shifted parameter vector. Thus, this formula is also known as the parameter shift rule \cite{schuld2020a}. Furthermore, the derivative of $\theta_j$ for the loss function can be obtained by the chain rule, 
\begin{eqnarray}\label{eq:14}
	\frac{\partial L_2 \left(\langle M \rangle \right)}{\partial \theta_j} = \frac{\partial L_2 \left(\langle M \rangle \right)}{\partial\langle M \rangle} \frac{\partial\langle M \rangle}{\partial \theta_j}.
\end{eqnarray}

Here we would like to remark that the estimation of the derivatives is performed in a quantum-classical hybrid way. Specifically, the terms $L_2(\vec{\theta} \pm \Delta \cdot e_j)$ in Equation~\ref{eq:12} or the terms $\langle M \rangle_{\vec{\theta} + \frac{\pi}{2} \cdot e_j}$ in Equation~\ref{eq:13} are evaluated by executing the quantum circuit, while the arithmetic with regard to these terms, including Equation~\ref{eq:14}, is done in the classical computer. Additionally, as can be seen above, estimating the gradients of the parameters is rather cost in quantum neural networks. Further works are required to clarify the limitation brought by the quantum gradient estimation and design a quantum backpropagation algorithm \cite{benedetti2019a}. 

Having the gradients of the parameters, the commonly used optimizers in classical neural networks can be used to update these parameters. Note that this step is performed in classical computers. We test the gradient descent and Adam optimizer. The numerical experiments show that both optimizers can work well; for comparison, Adam can give a $32\%$ improvement in the training speed, while has a slight decrease in accuracy.

\section{Experimental results}\label{sec:4}
In order to verify the performance of the present QRNN models, we evaluate our QRNNs on three different kinds of classical sequential data, i.e., meteorological indicators, stock price, and natural language. In the numerical experiments, both two architectures, i.e., pQRNN and sQRNN are evaluated. The difference method of estimating the gradients and the gradient descent optimizer are used to update the parameters. The algorithms are implemented based on the pyQPanda quantum programming framework \cite{2022c}. The following experiments show that the present QRNN models achieve promising performance on the three totally different sequential data. 

\subsection{Meteorological indicators}
The meteorological data contains five indicators, which are the atmospheric pressure, maximum temperature, minimum temperature, relative humidity, and wind speed. The sequence of each indicator contains $500$ elements representing the values of the indicator on $500$ days. The task is to train the QRNN model to be capable of predicting the value of each indicator on the eighth day using the values of preceding seven days.  

The circuits of the pQRNN and sQRNN used here are the ones as shown in Fig.~\ref{fig:7} and~\ref{fig:8}. Specifically, the number of qubits used in Reg. D (also in Reg. H) is $3$; the number of QRB is seven to embed the seven days’ data and the output of the seventh QRB is taken as the prediction of data on the eighth day. The 500 elements of each indicator are divided into $300$ for training and $200$ for test. The hyperparameter of learning rate is $0.03$. In addition, to be as a reference, a classical RNN is constructed with the structure as shown in Fig.~\ref{fig:1}, where the number of recurrent block is also seven for fair comparison with QRNN.

The prediction accuracy of each indicator is evaluated using the following formula,
\begin{eqnarray}
	\begin{aligned}
	Accuarcy &=\left( 1- \sqrt{\frac{1}{N}\sum_{i=1}^N E_i^2}\right) \times 100\%,  \\
	E_i &= \frac{actual - predicted}{actual},
	\end{aligned}
\end{eqnarray}
where $E_i$ is the relative error of each predicted value (i.e., the data of eighth day) and $N$ is the total number of predictions (i.e., the number of samples for test). 

The prediction accuracy of the two QRNN models as well as the classical RNN are shown in Table~\ref{tab:1}. The first remarkable conclusion obtained from the table is that the sQRNN model can achieve a similar accuracy as pQRNN. This is impressive because sQRNN can be implemented efficiently on the NISQ devices, and together with the great performance implies that sQRNN has the potential to find useful applications on near-term quantum devices.

The second remarkable conclusion is that both two QRNN models are capable of predicting the five indicators with higher accuracy than the classical RNN. In particular, for the indicator of wind speed, the performances of QRNN are much better than the classical RNN. The relative variation of wind speed is much acuter that other indicators. Hence, it implies that QRNNs should possess a better capability of predicting the trend of acute variation. 

\begin{table}[ht]
\begin{center}
	\caption{Accuracy of the two QRNN models and the classical RNN on the sequential data of meteorological indicators.} \label{tab:1}
	\begin{tabular}{cccccc}
		\toprule
		\multirow{2}{*}{\textbf{Meteorological indicators}}  & \multicolumn{3}{c}{\textbf{Prediction accuary}}  \\ \cmidrule(lr){2-4}
		{} &\textbf{pQRNN} & \textbf{sQRNN} &\textbf{RNN} \\ 
		\midrule
		Atmospheric Pressure & 99.91\%  & 99.87\% & 99.83\% \\  
		Minimum Temperature  & 96.96\% & 93.09\%  & 92.65\% \\
		Maximum Temperature  & 97.68\% & 90.15\% &91.29\% \\
		Relative Humidity  & 98.51\%  & 96.19\%  &92.99\% \\
		Wind Speed  & 90.13\%  & 87.96\% & 70.38\% \\
		\bottomrule
	\end{tabular}
\end{center}
\end{table}

\begin{figure}[ht]
	\centering
	\includegraphics[scale=0.45]{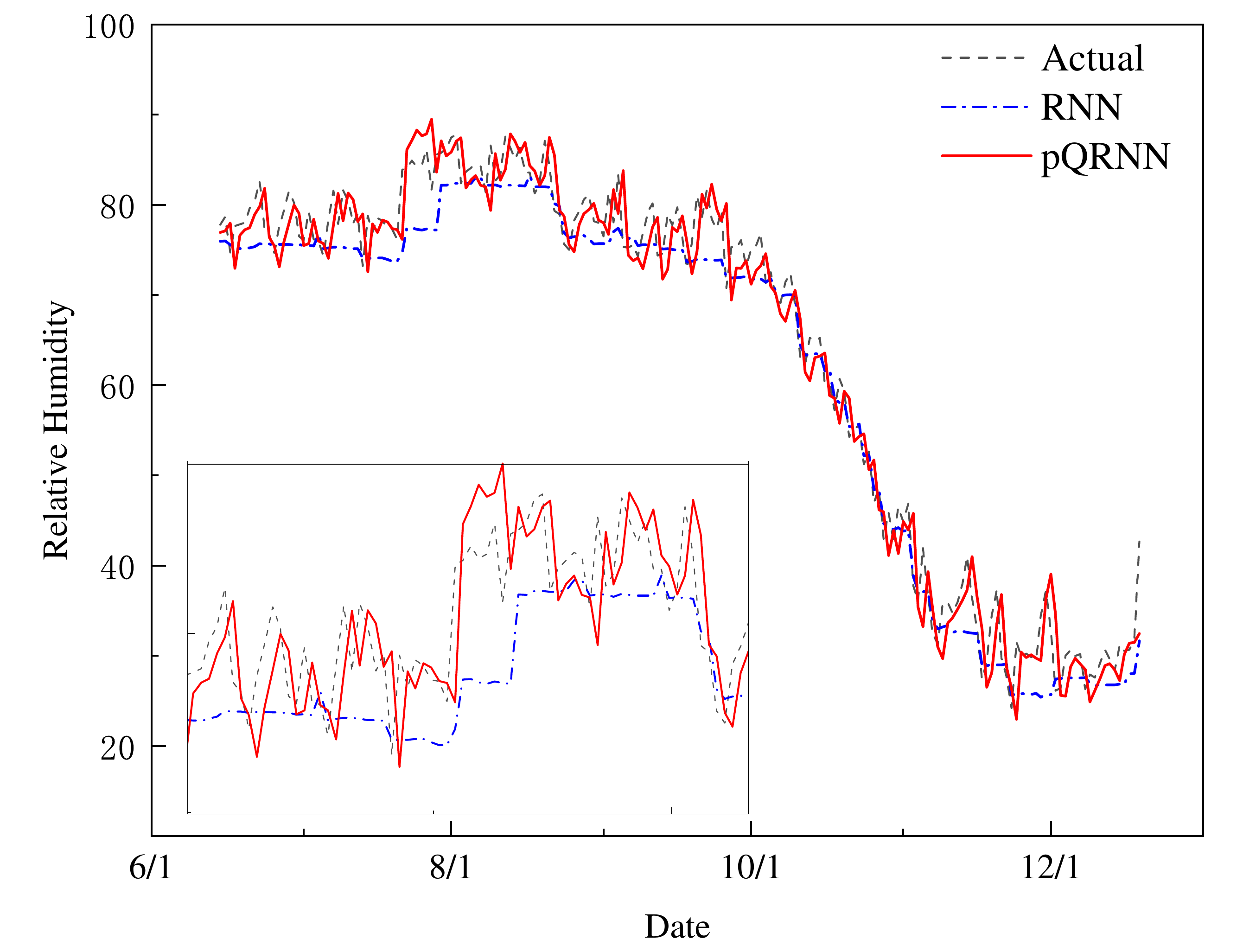}
	\caption{Curves of the indicator of relative humidity predicted by pQRNN and RNN.}
	\label{fig:9}
\end{figure}

Let us make a further comparison between QRNN and the classical RNN. In QRNN, the number of parameters required to learn are $30$, while in RNN it is 49. That is, QRNNs use less parameters but get a higher prediction accuracy. This should result from the fact that PQCs have stronger expressibility than the classical networks \cite{du2022c,du2020}. In other words, with the similar number of parameters, the function space achieved by PQC is exponentially larger than that of classical networks. Furthermore, we find that the QRNNs are capable of providing better predictions of the variation details of the temporal data than the classical RNN. Fig.~\ref{fig:9} shows the curves of relative humidity indicator predicted by pQRNN and RNN. As can be seen from the figure, Details of the fluctuation presented in the actual sequence data are learned by QRNN, while classical RNN has an effect of smoothing the fluctuation. This, together with the above phenomena that QRNNs perform much better when predicting the wind speed whose variation is much acuter, shows that our QRNNs can capture the changing details existing in the temporal sequence data more efficiently. More predicted curves of other indicators are presented in~\ref{append:A}. 

In order to verify the flexibility of the circuit structure of QRNN, we test the prediction accuracy of pQRNN with different number of qubits. Specifically, the number of qubits used in Reg. D (also in Reg. H) is set to $4$, $6$, and $8$. The results are shown in Table~\ref{tab:2}. As can be seen from the table, generally higher prediction accuracy can be obtained when using more qubits in the QRNN circuit. However, there exits the phenomena of accuracy saturation; that is, with the increase of the number of qubits, the improvement of accuracy would become small. 

\begin{table}[ht]
\begin{center}
	\caption{Prediction accuracy of the pQRNN with 4, 6 and 8 qubits in Reg. D (and Reg. H).}\label{tab:2}
	\begin{tabular}{cccc}
		\toprule
		\multirow{2}{*}{\textbf{Meteorological indicators}}  & \multicolumn{3}{c}{\textbf{Prediction accuary}}  \\ \cmidrule(lr){2-4}
		{} &\textbf{4-qubits} & \textbf{6-qubits} &\textbf{8-qubits} \\ 
		\midrule
		Atmospheric Pressure & 99.94\% & 99.98\% & 99.99\% \\  
		Minimum Temperature  & 94.89\% & 96.96\% & 97.07\% \\
		Maximum Temperature  & 97.05\% & 97.68\% & 98.08\% \\
		Relative Humidity  & 98.32\% & 98.51\% & 98.30\% \\
		Wind Speed  & 87.57\% & 90.13\%  & 91.12\% \\
		\bottomrule
	\end{tabular}
\end{center}
\end{table}

\subsection{Stock price}
The second task is to predict the variation of stock price, including the opening price, maximum price, minimum price, closing price, and volume of a stock. The sequence data of each component of stock price contains $180$ elements representing the price of $180$ days. The task is to use the first seven days’ data to predict the price on the eighth day. The circuits of the pQRNN and sQRNN used here as well as the classical RNN are the same as those used in predicting meteorological indicators. The $180$ elements of each component are divided into $100$ for training and $80$ for test. The hyperparameter of learning rate is $0.03$.

The prediction accuracy of the two QRNN models as well as the classical RNN are shown in Table~\ref{tab:3}. Almost the same conclusions can be obtained: (1) the sQRNN model can achieve a similar accuracy as pQRNN; (2) both two QRNN models are capable of predicting the five components of stock price with higher accuracy than the classical RNN. Therefore, the advantages of QRNN against classical RNN can be embodied on different kinds of learning data. The predicted curves of five components of stock price are presented in~\ref{append:B}.

\begin{table}[ht]
\begin{center}
	\caption{Prediction accuracy of the two QRNN models and the classical RNN on the sequential data of stock price.}\label{tab:3}
	\begin{tabular}{cccccc}
		\toprule
		\multirow{2}{*}{\textbf{Stock indicators}}  & \multicolumn{3}{c}{\textbf{Prediction accuary}}  \\ \cmidrule(lr){2-4}
		{} &\textbf{pQRNN} & \textbf{sQRNN} &\textbf{RNN} \\ 
		\midrule
		Opening price & 98.69\%  & 97.29\%  & 95.83\% \\  
		Highest price  & 98.83\%  & 97.99\%  & 96.68\% \\
		Lowest price  & 98.82\% & 97.89\%  & 96.20\% \\
		Closing price  & 99.08\% & 97.61\% &96.11\% \\
		Volume  & 90.36\% & 87.13\% &73.73\% \\
		\bottomrule
	\end{tabular}
\end{center}
\end{table}

\subsection{Text categorization}
The above two tasks are regression problems, while the third task is a classification problem. In order to verify the feasibility of QRNNs to perform nature language processing, we use the MC (meaning classification) task to test our QRNNs. In the MC task, there contains $130$ sentences and each sentence has $3$ or $4$ words. Half of the sentences are related to food and half to information technology (IT). Hence, MC is a binary classification task that categorizes a sentence as food or IT. There are totally $17$ words in MC and part of the vocabulary is in common between the two classes, so the task is not trivial \cite{lorenz2021b}. 

The circuits of the pQRNN and sQRNN used here are the same as above. The $130$ sentences are divided into $100$ for training and $30$ for test. The hyperparameter of learning rate is $0.01$.

The classification accuracy of the two QRNN models as well as two state-of-the-art QNN models for natural language learning are shown in Table~\ref{tab:4}. As can be seen from the table, both two QRNN models can achieve an accuracy of $100\%$, which is much better than that of the syntactic analysis-based quantum model \cite{lorenz2021b}. On the other hand, the accuracy of QSANN (quantum self-attentive neural networks) proposed in Ref. \cite{li2022b} also achieves $100\%$, but the computational cost of QSANN is much heavier than the present QRNNs. In addition, QSANN is a quantum-classical hybrid model, where the queries, keys, and values are implemented by the PQCs and the self-attention coefficients are calculated in the classical resources. QSANN uses $12$ quantum parameters to encode each word into the circuit, while in our QRNNs, only one parameter is enough.

\begin{table}[ht]
\begin{center}
	\caption{Classificaion accuracy of the two QRNN models and two state-of-the-art QNN models for sequential learning on the MC task.}\label{tab:4}
\begin{tabular}{cccc}
	\toprule
	\multicolumn{4}{c}{\textbf{Prediction accuracy}} \\
	\cmidrule(lr){1-4}
	\textbf{pQRNN} & \textbf{sQRNN} & \textbf{QSANN}~\cite{li2022b} &\textbf{DisCoCat}~\cite{lorenz2021b}\\
	100\% & 100\% & 100\% & 79.8\% \\
	\bottomrule
\end{tabular}
\end{center}
\end{table}

\section{Conclusion}\label{sec:5}
In the present work, we develop a hardware-efficient way of constructing the quantum recurrent blocks, and by stacking the blocks in a staggered way we obtain the staggered QRNN model that can  be implemented efficiently on quantum devices with much lower requirement of the coherent time. The efficiency of the present QRNN models are verified using three different kinds of classical sequential data, i.e., the meteorological indicators, stock price, and text categorization. The numerical experiments show that our QRNN models  have a much better performance in prediction (classification) accuracy than the classical RNN and state-of-the-art QNN models for sequential learning, and can capture the acute variation trend existing in the temporal sequence data. In one word, the present QRNNs possess a simple and well-designed structure, and show great performance on both regression and classification problems.  

We think the present work is a significative start of studying the quantum recurrent neural networks for classical sequential data. The present QRNN model would be taken as a candidate of canonical QRNN model to study the possible near-term applications of quantum deep learning. The interesting future work includes (1) optimizing the present QRNNs further, e.g., applying different data encoding methods; (2) characterizing the trainability of the QRNNs, i.e., the properties of barren plateau and landscape of cost functions; (3) expanding the present models to be a deep recurrent neural network, reminiscent to the LSTM and GRU networks in deep learning. 

\section*{Acknowledgments}
The present work is supported by the Natural Science Foundation of Shandong Province of China (ZR2021ZD19) and the National Natural Science Foundation of China (Grant No. 12005212). 

We are grateful to the support from Big Data Center of Marine Advanced Research Institute of Ocean University of China. We also thank the technical team from the Origin Quantum Computing Technology Co., Ltd, in Hefei for their professional services.


\bibliographystyle{elsarticle-num-names} 
\bibliography{Quantum_Recurrent_neural_network_for_sequential_learning}

\newpage
\appendix

\section{Curves of the meteorological indicators predicted by pQRNN and sQRNN}\label{append:A}
\subsection{Curves of the meteorological indicators predicted by pQRNN}
	
\begin{figure}[ht]
	\centering
	\subfigure[]{
		\includegraphics[scale=0.15]{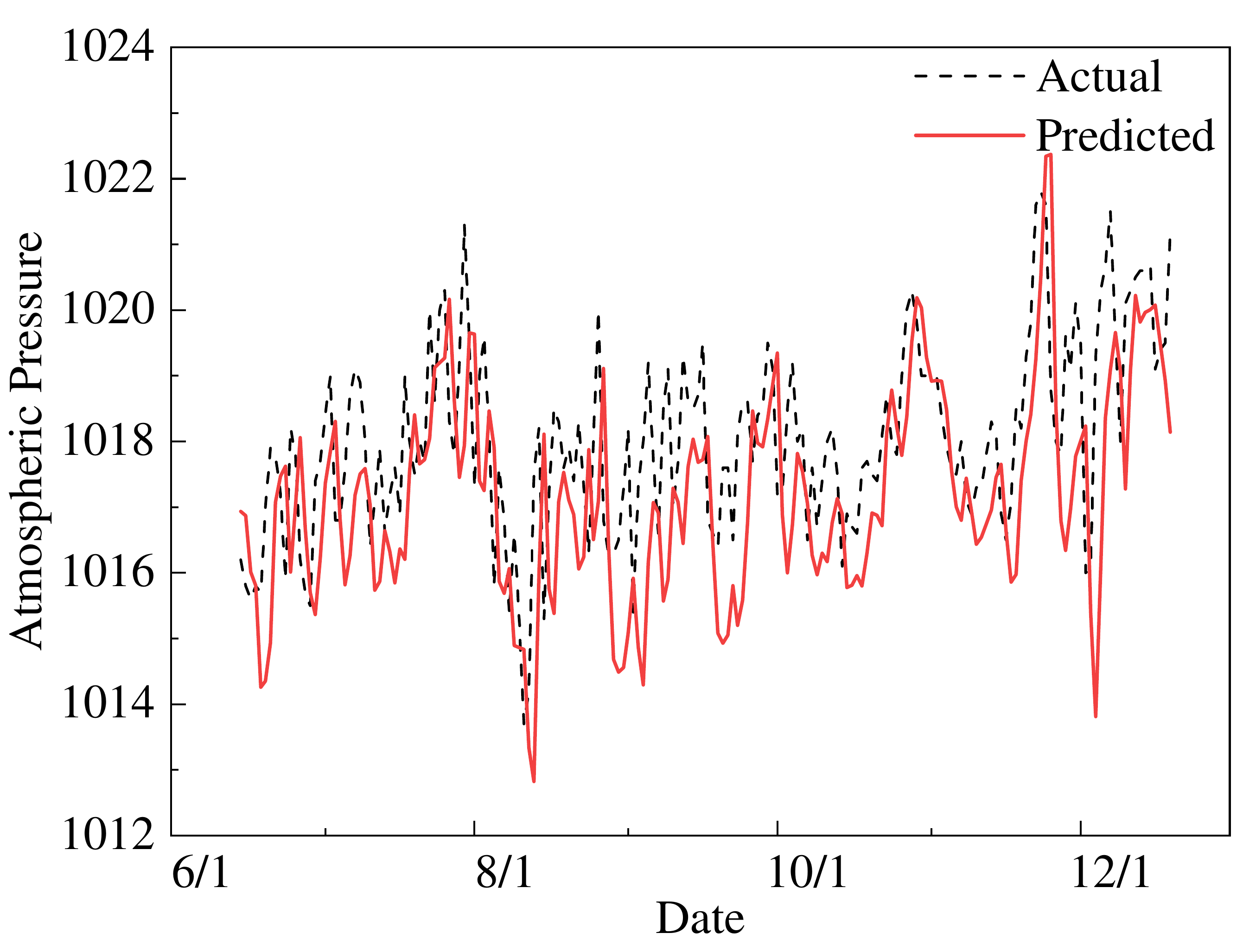} 
	}
	\quad
	\subfigure[]{
		\includegraphics[scale=0.15]{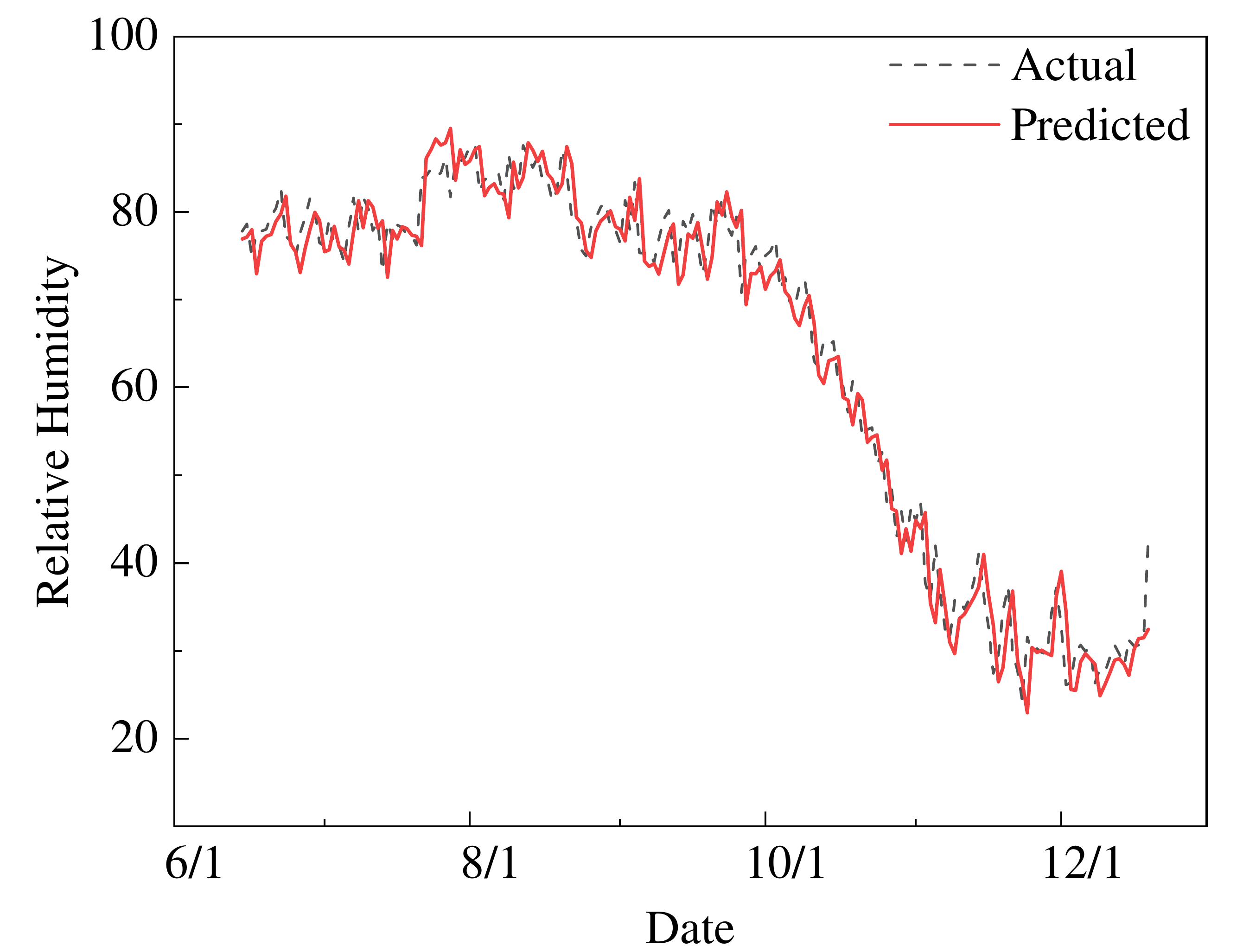} 
	}
	\quad
	\subfigure[]{
		\includegraphics[scale=0.15]{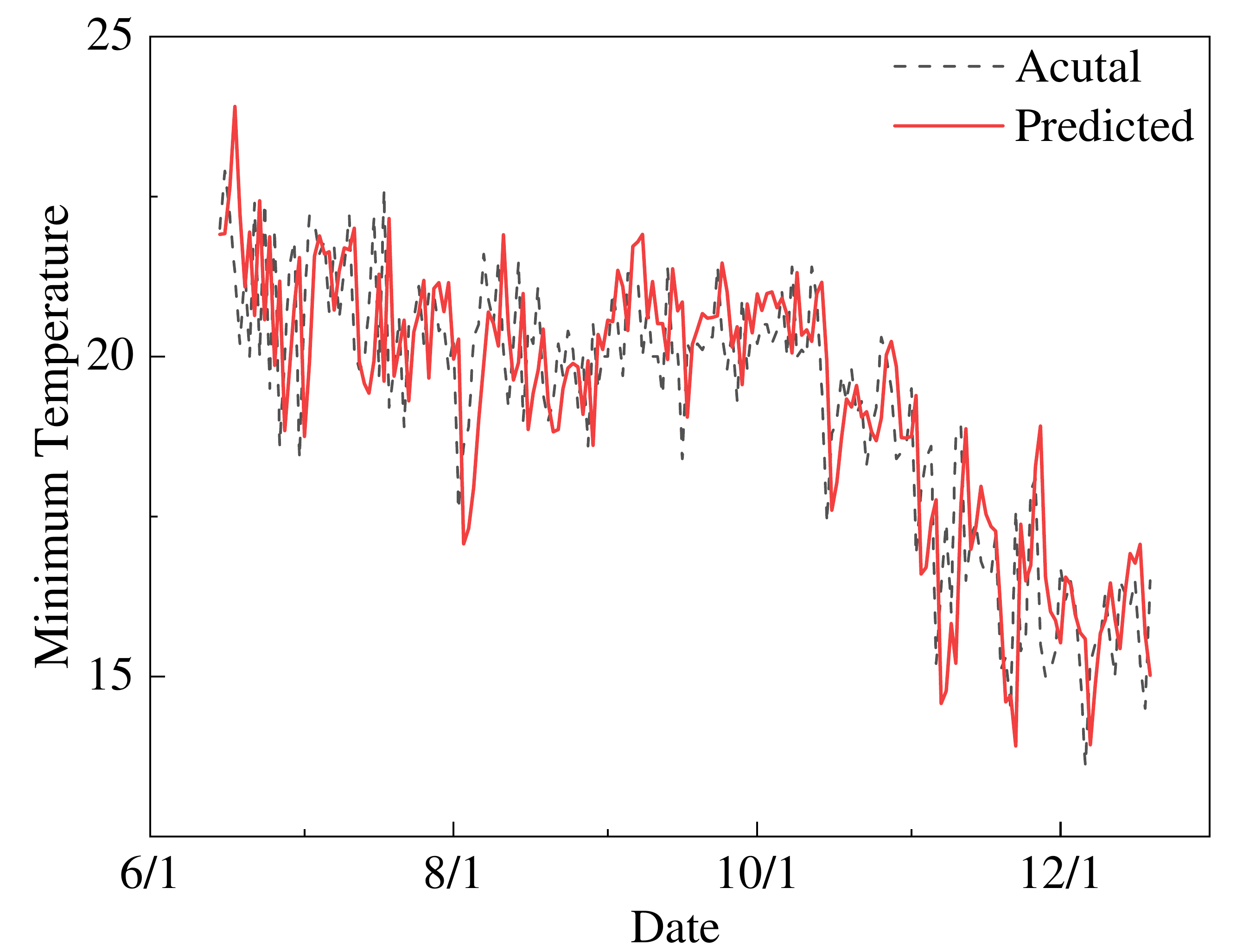} 
	}
	\quad
	\subfigure[]{
		\includegraphics[scale=0.15]{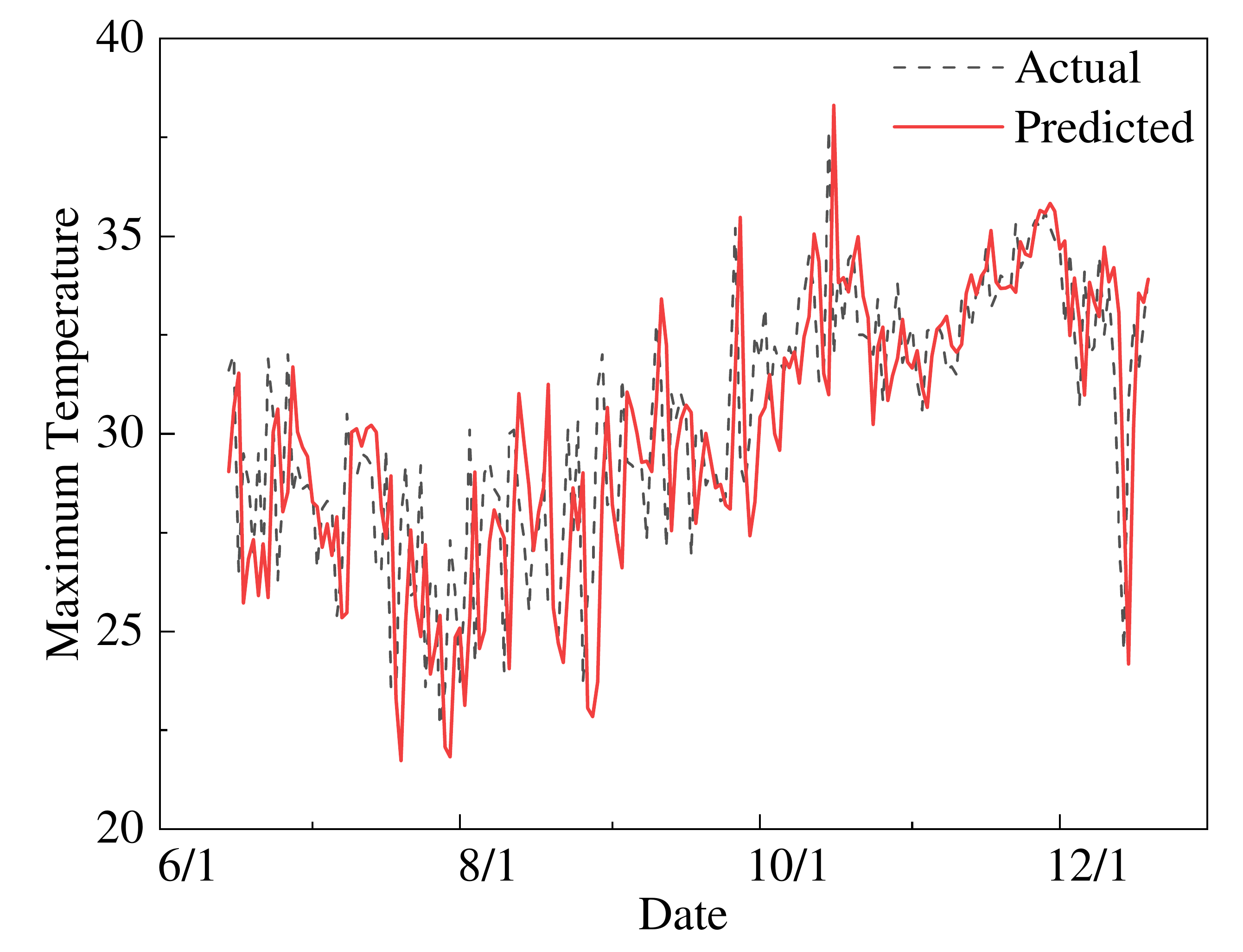} 
	}
	\quad
	\subfigure[]{
		\includegraphics[scale=0.15]{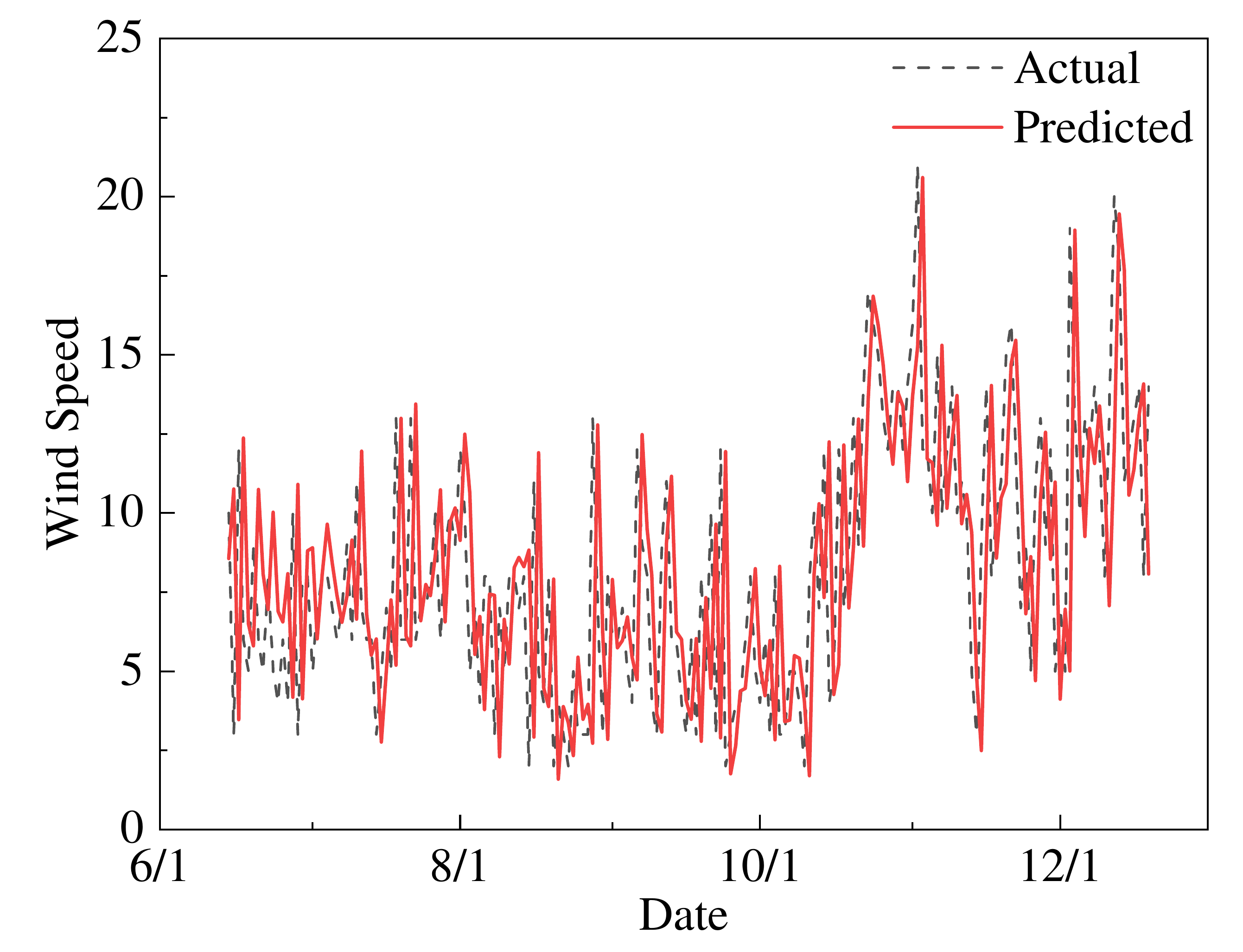} 
	}
	\caption{Curves of the meteorological indicators predicted by pQRNN: (a) atmospheric pressure, (b) relative humidity, (c) minimum temperature, (d) maximum temperature, and (e) wind speed.}
\end{figure}

\newpage
\subsection{Curves of the meteorological indicators predicted by sQRNN}

\begin{figure}[h]
	\centering
	\subfigure[]{
		\includegraphics[scale=0.15]{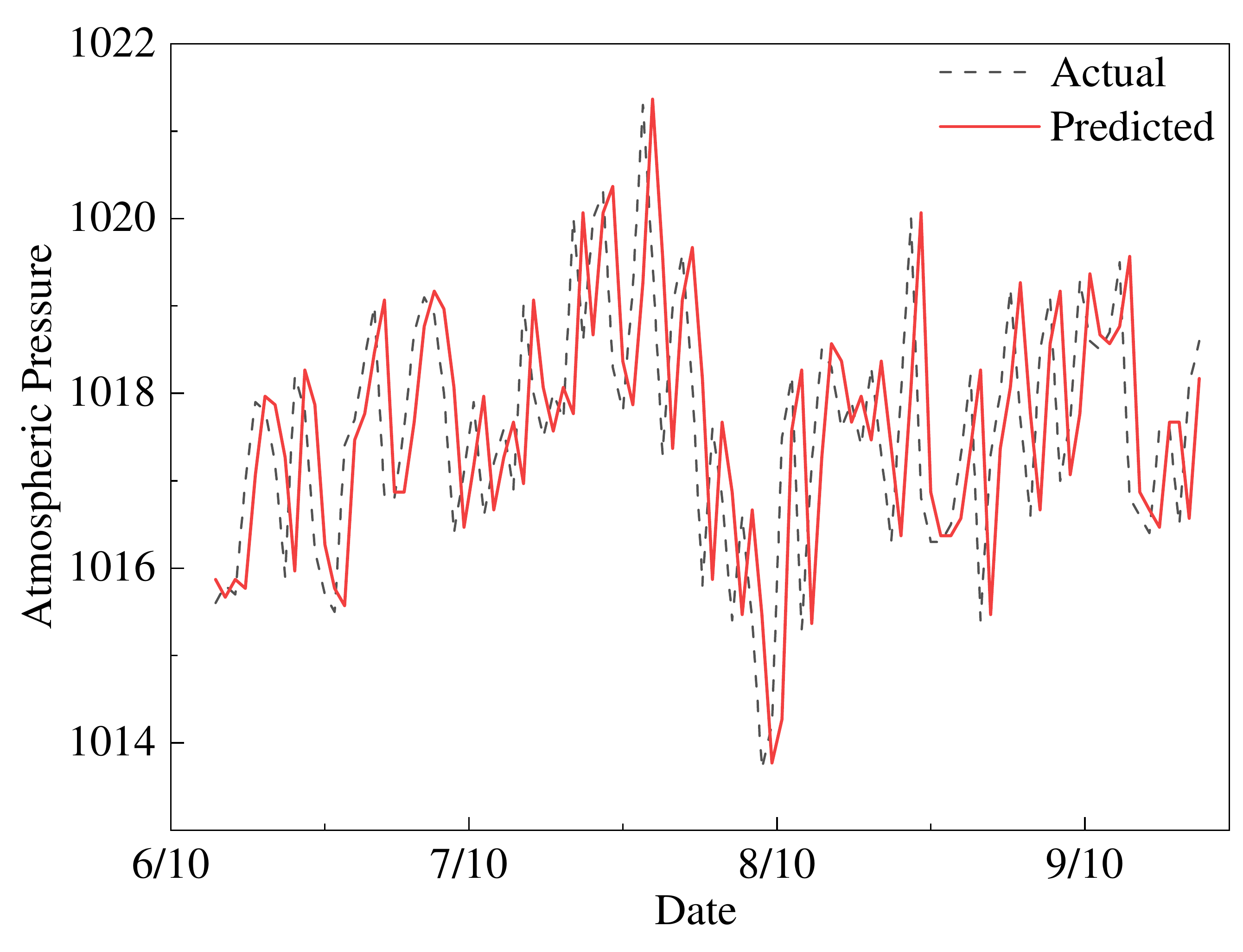} 
	}
	\quad
	\subfigure[]{
		\includegraphics[scale=0.15]{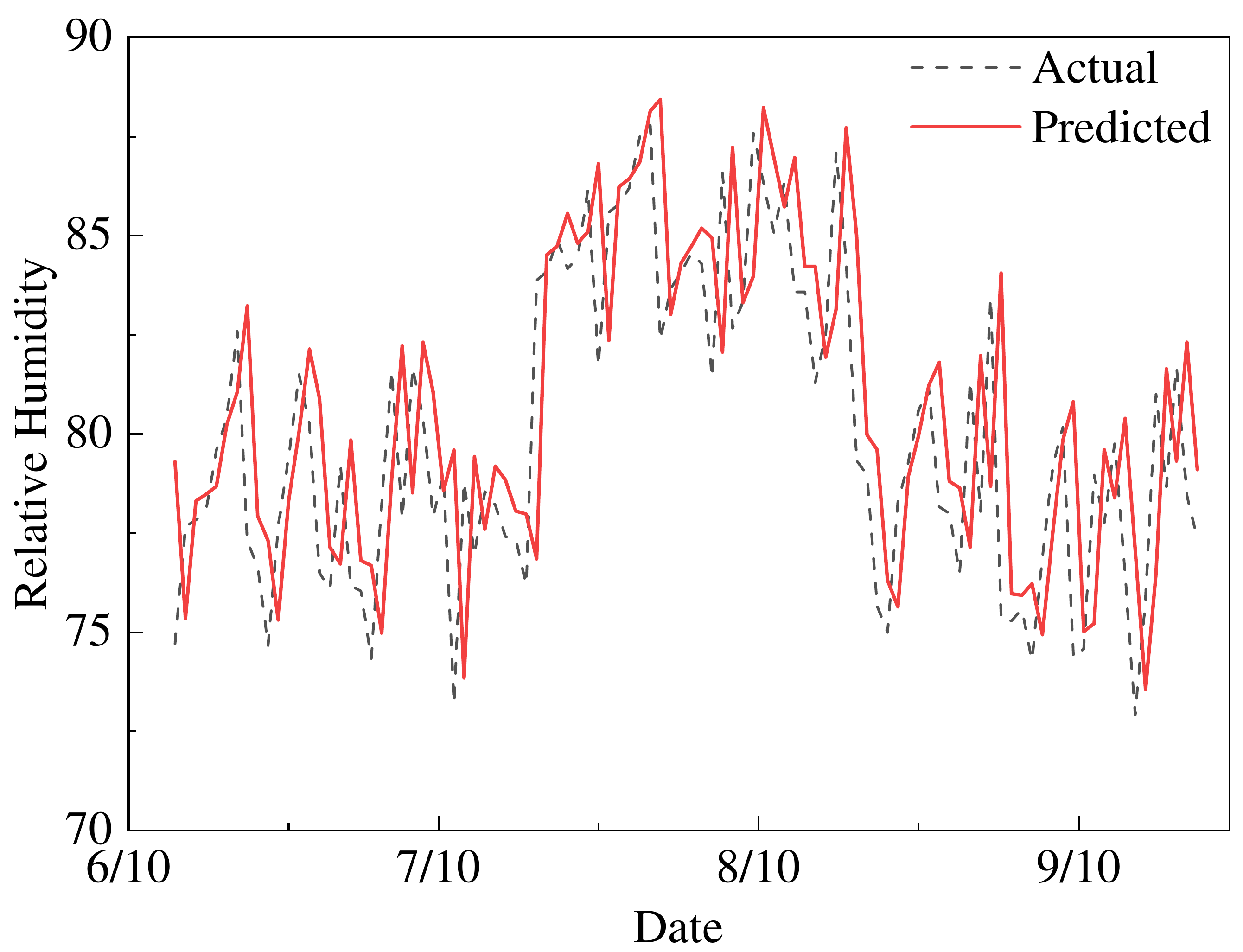} 
	}
	\quad
	\subfigure[]{
		\includegraphics[scale=0.15]{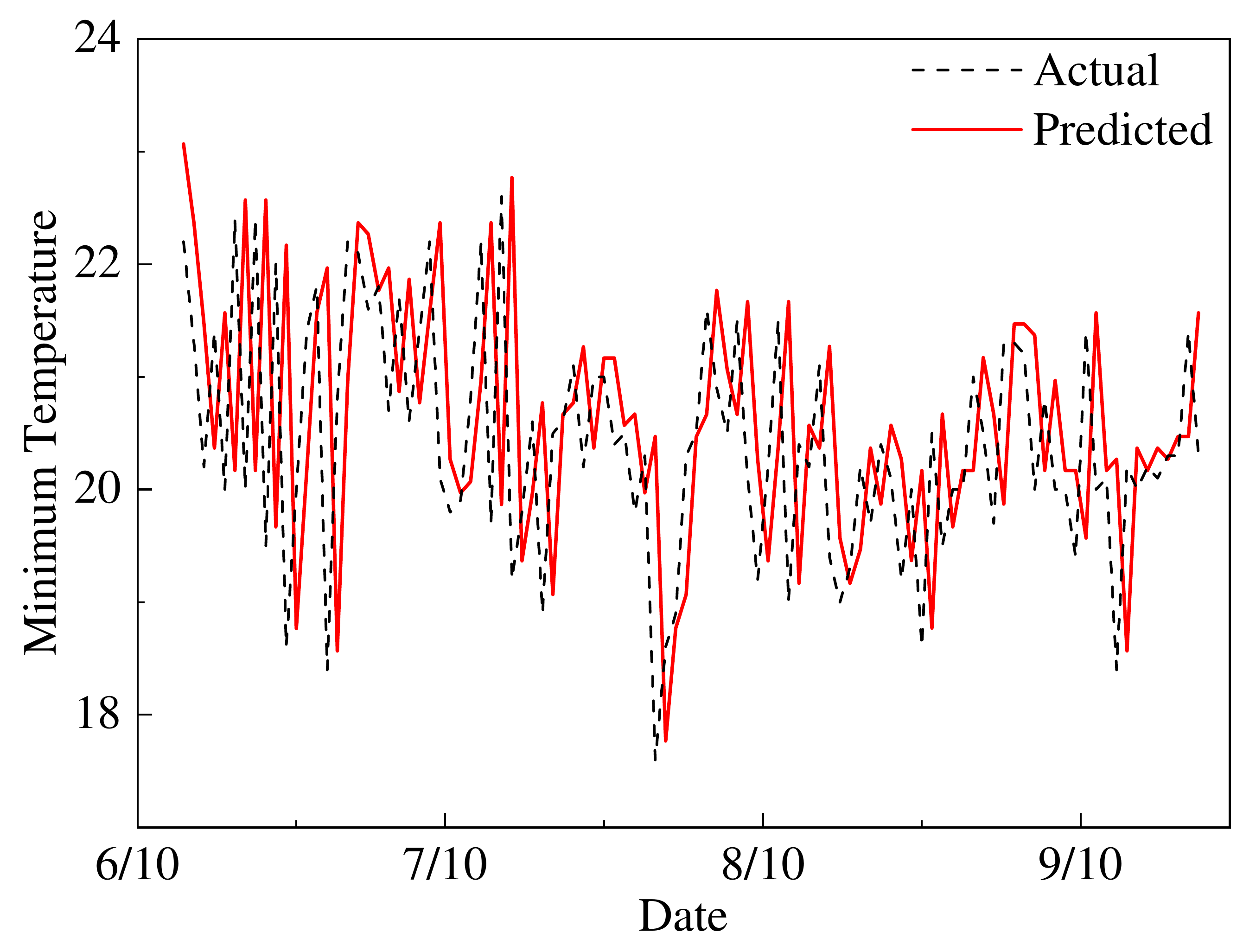} 
	}
	\quad
	\subfigure[]{
		\includegraphics[scale=0.15]{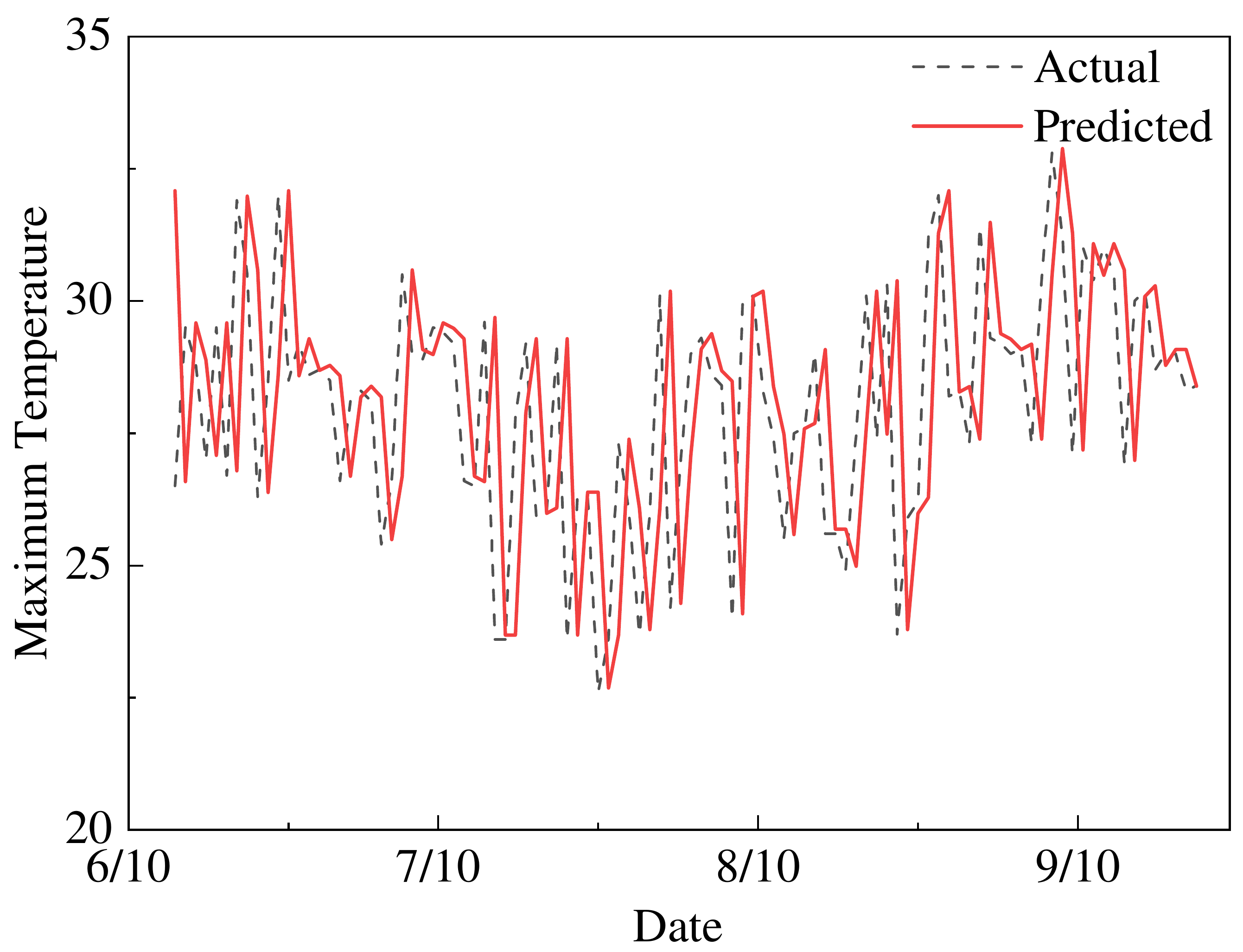} 
	}
	\quad
	\subfigure[]{
		\includegraphics[scale=0.15]{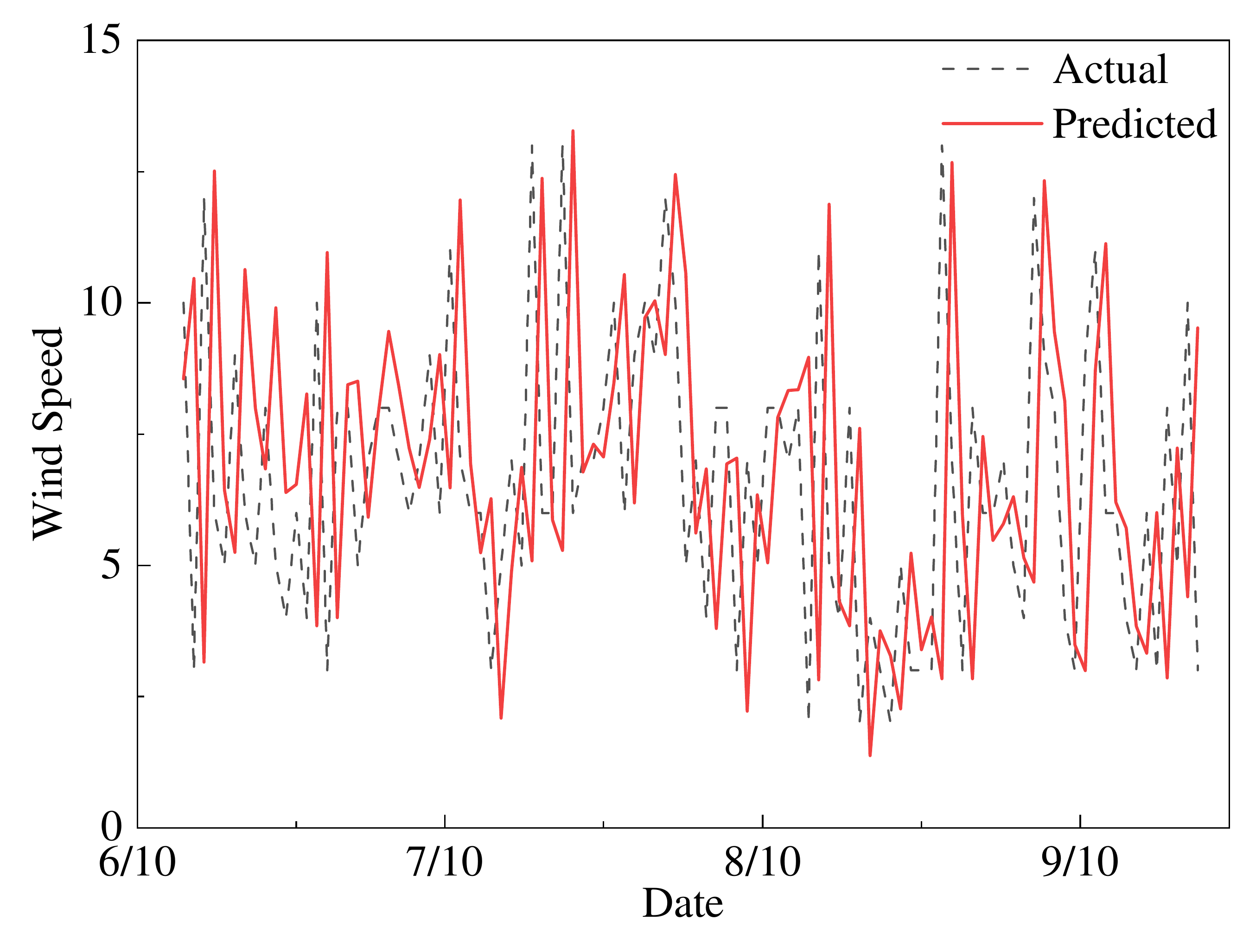} 
	}
	\caption{Curves of the meteorological indicators predicted by sQRNN: (a) atmospheric pressure, (b) relative humidity, (c) minimum temperature, (d) maximum temperature, and (e) wind speed.}
\end{figure}

\newpage
\section{Curves of the stock price predicted by pQRNN and sQRNN}\label{append:B}

\subsection{Curves of the stock price predicted by pQRNN}

\begin{figure}[ht]
	\centering
	\subfigure[]{
		\includegraphics[scale=0.15]{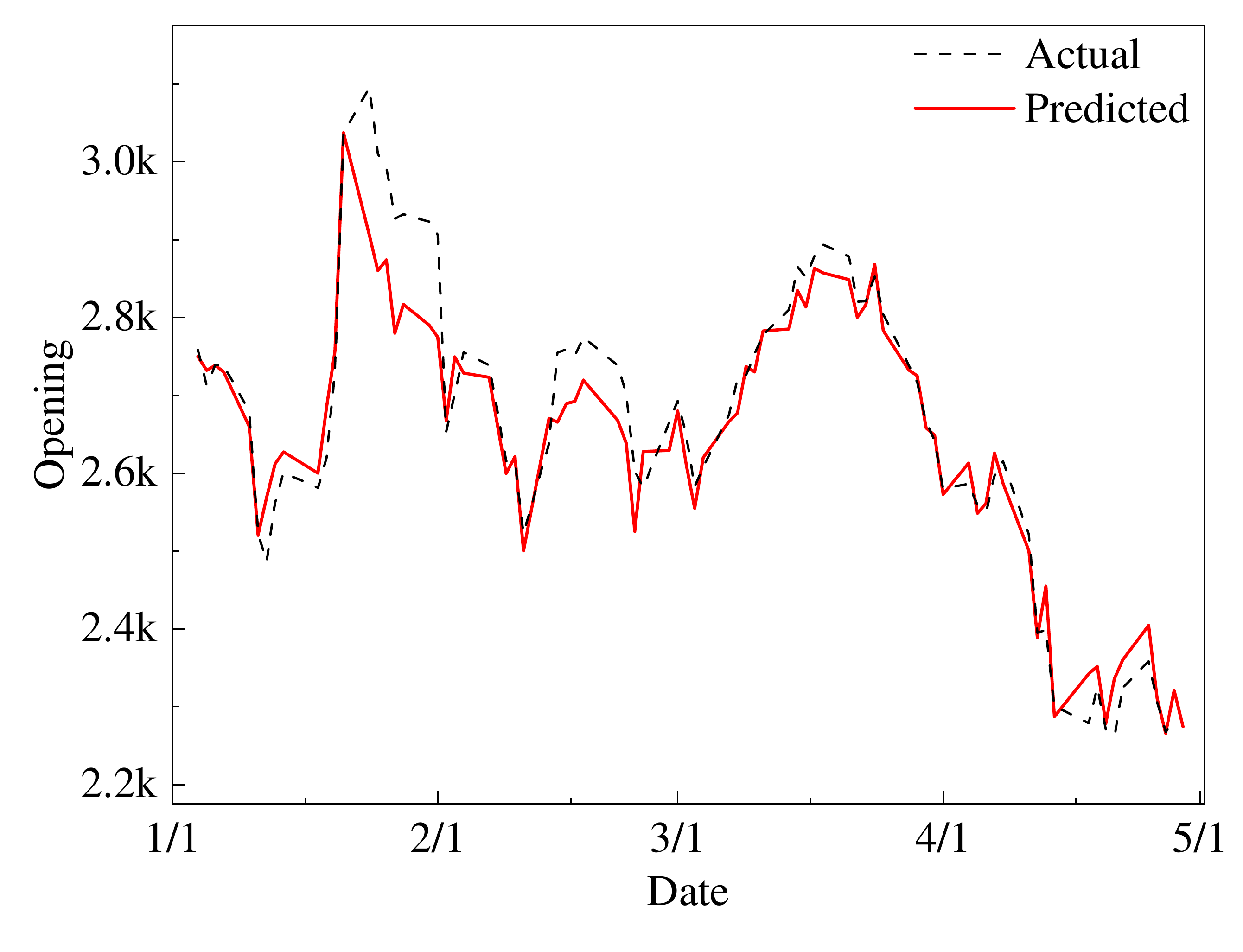} 
	}
	\quad
	\subfigure[]{
		\includegraphics[scale=0.15]{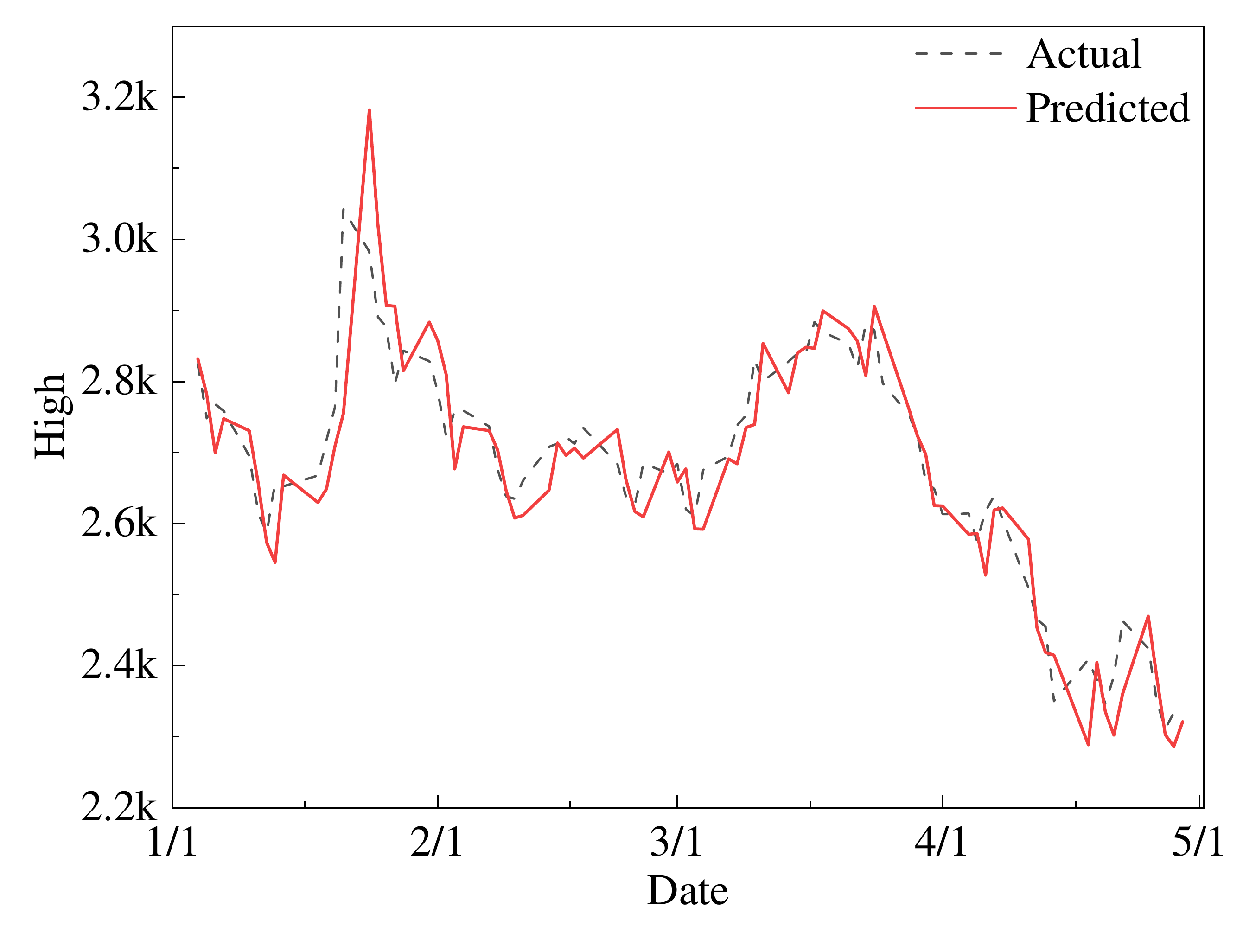} 
	}
	\quad
	\subfigure[]{
		\includegraphics[scale=0.15]{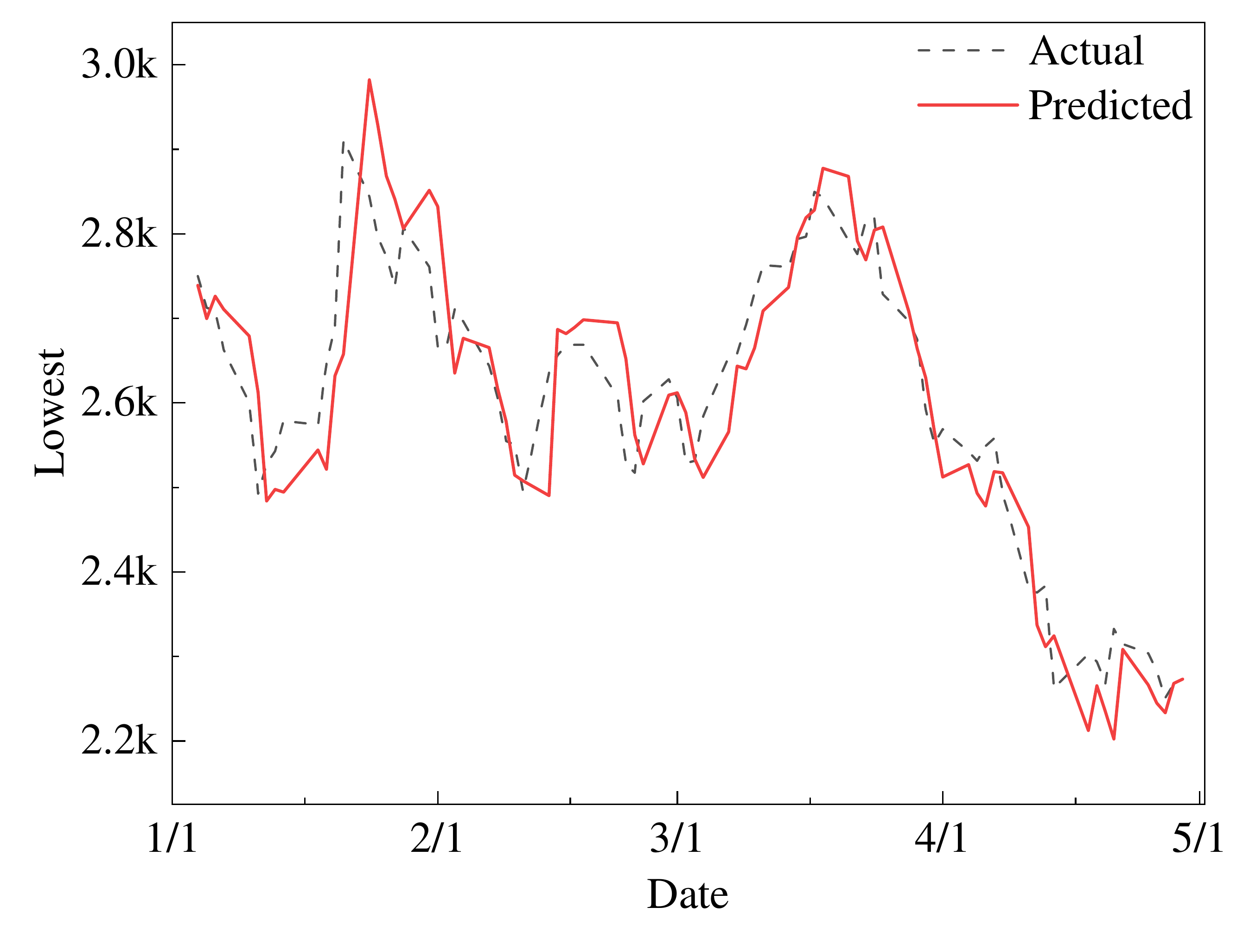} 
	}
	\quad
	\subfigure[]{
		\includegraphics[scale=0.15]{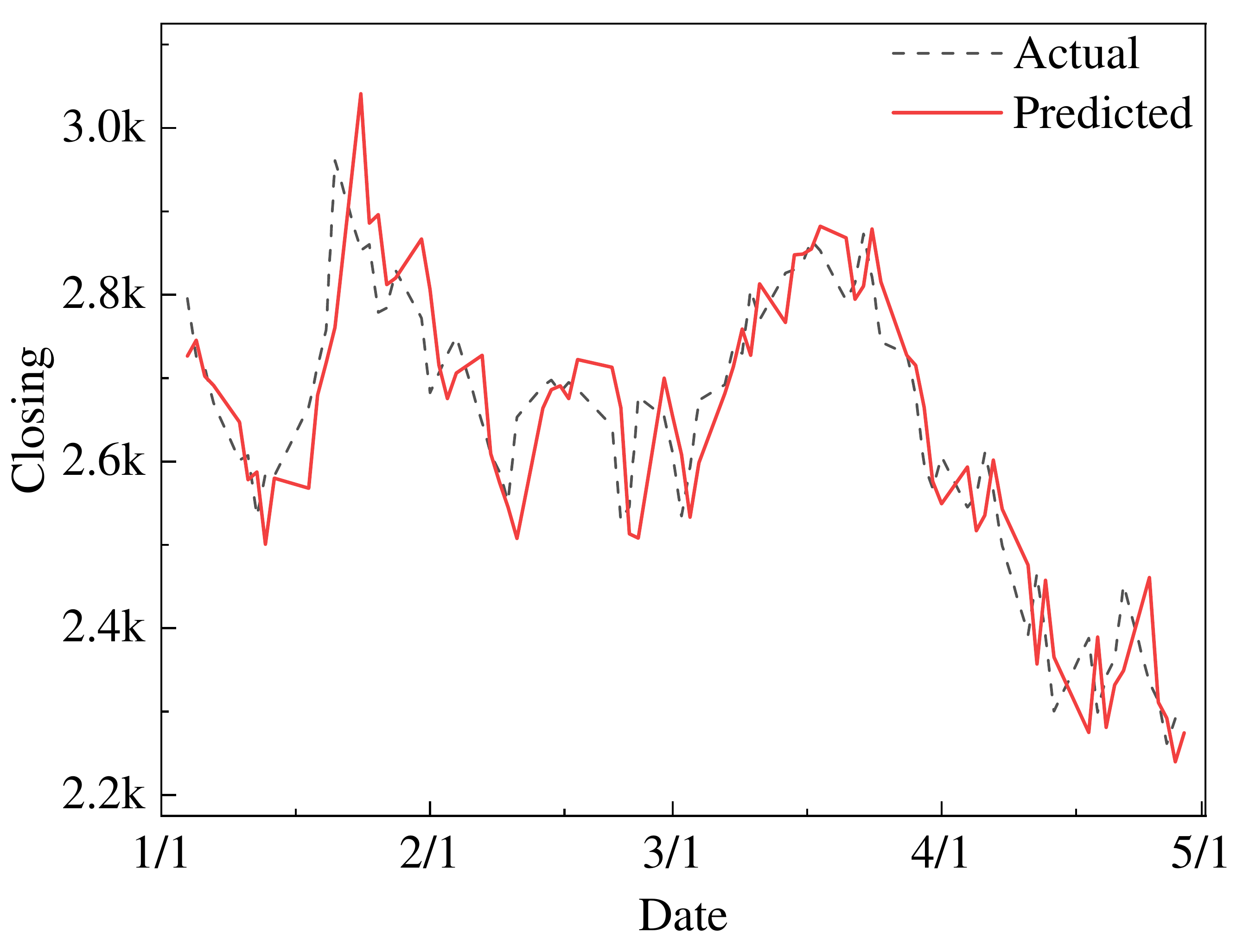} 
	}
	\quad
	\subfigure[]{
		\includegraphics[scale=0.15]{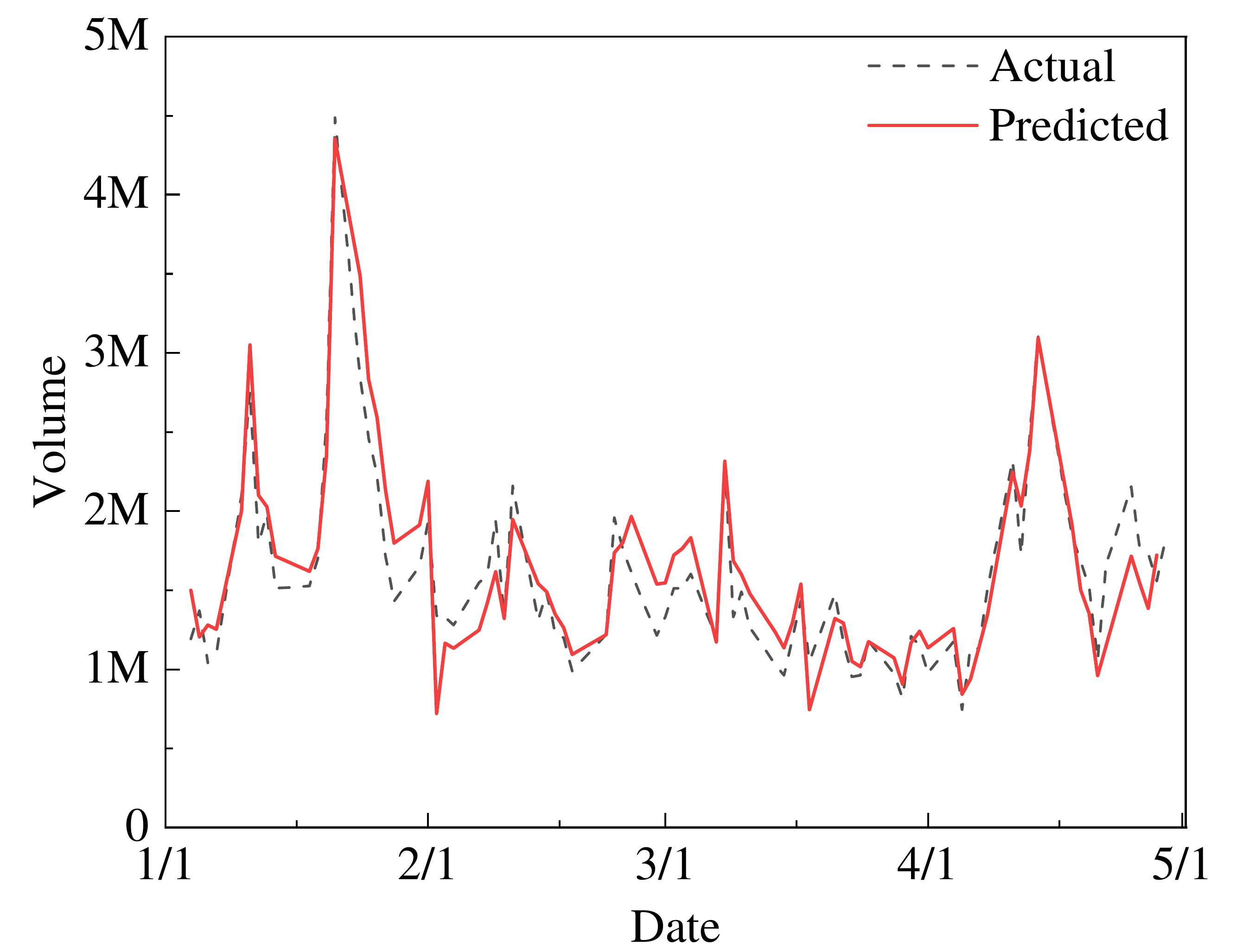} 
	}
	\caption{Curves of the stock price predicted by pQRNN: (a) the opening price, (b) the highest price, (c) the lowest price, (d) the closing price, and (e) the volume. }
\end{figure}

\newpage
\subsection{Curves of the stock price predicted by sQRNN}

\begin{figure}[ht]
	\centering
	\subfigure[]{
		\includegraphics[scale=0.15]{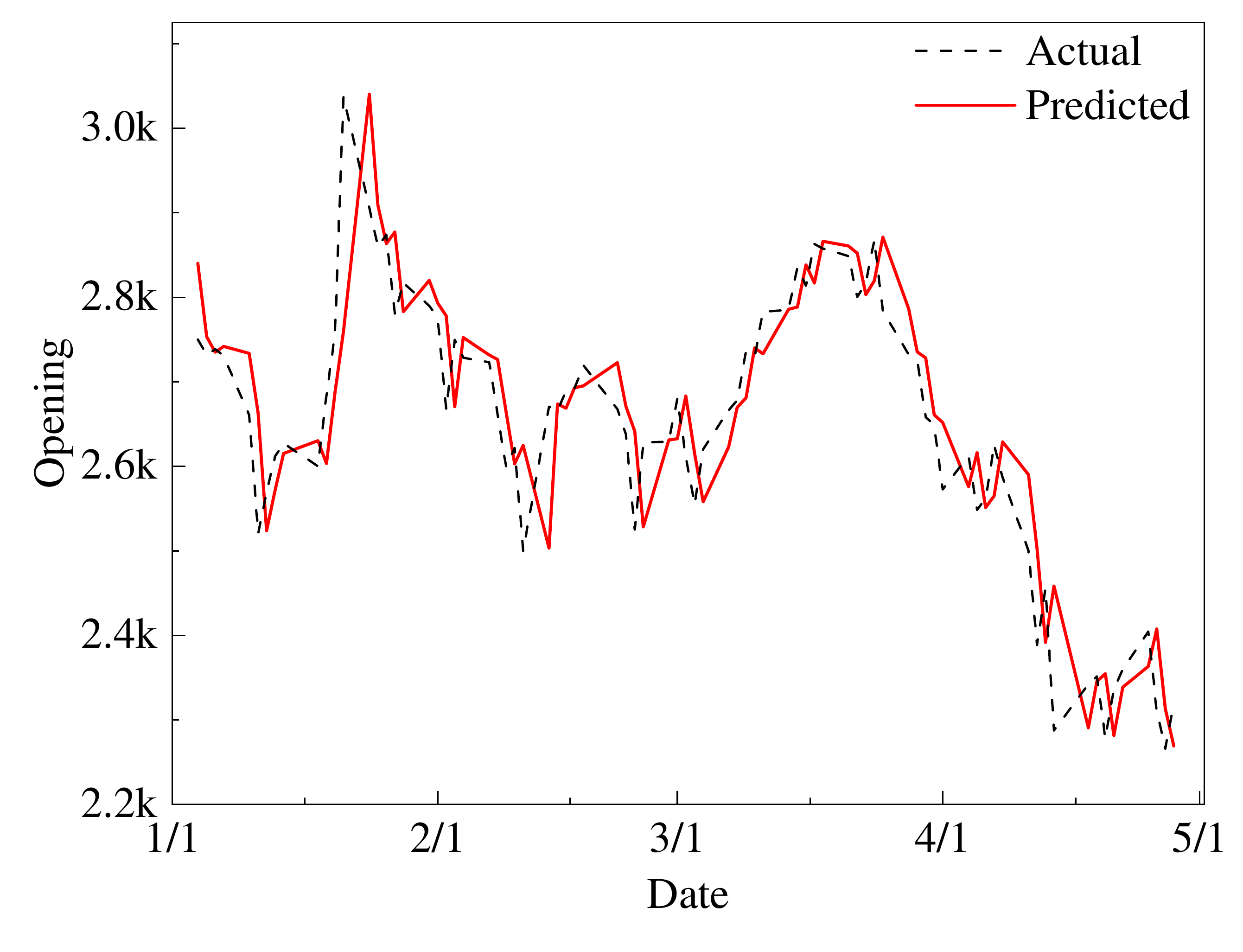} 
	}
	\quad
	\subfigure[]{
		\includegraphics[scale=0.15]{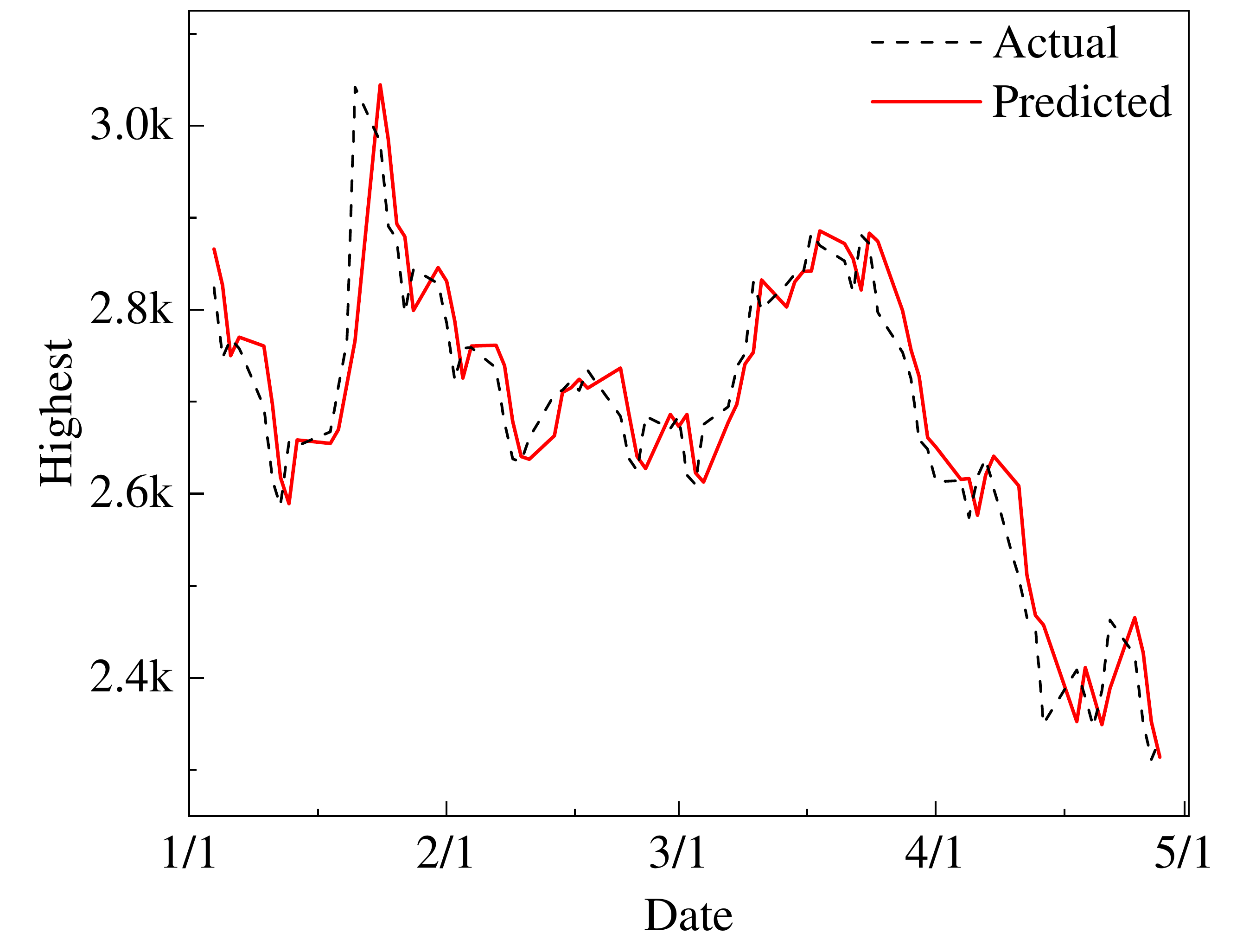} 
	}
	\quad
	\subfigure[]{
		\includegraphics[scale=0.15]{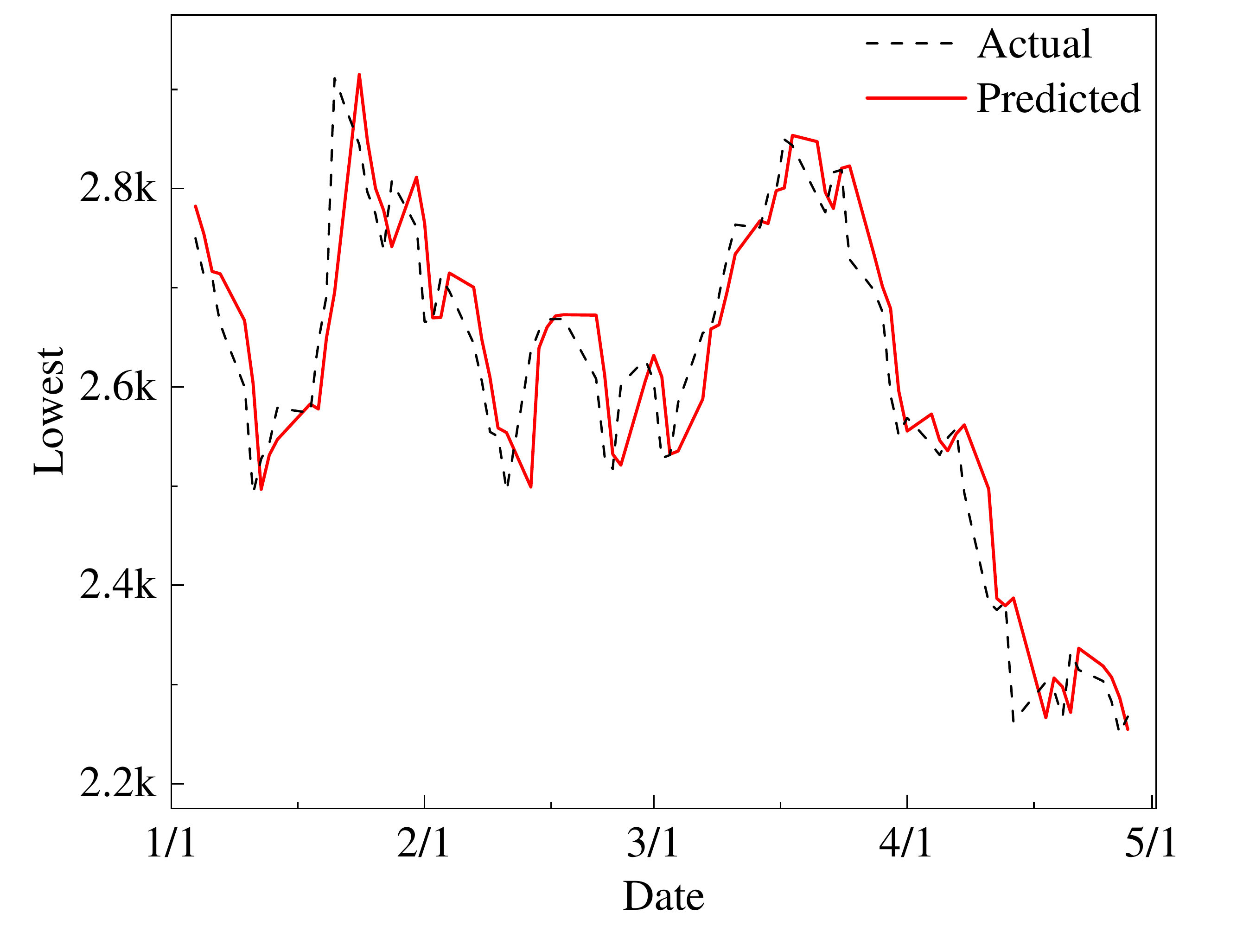} 
	}
	\quad
	\subfigure[]{
		\includegraphics[scale=0.15]{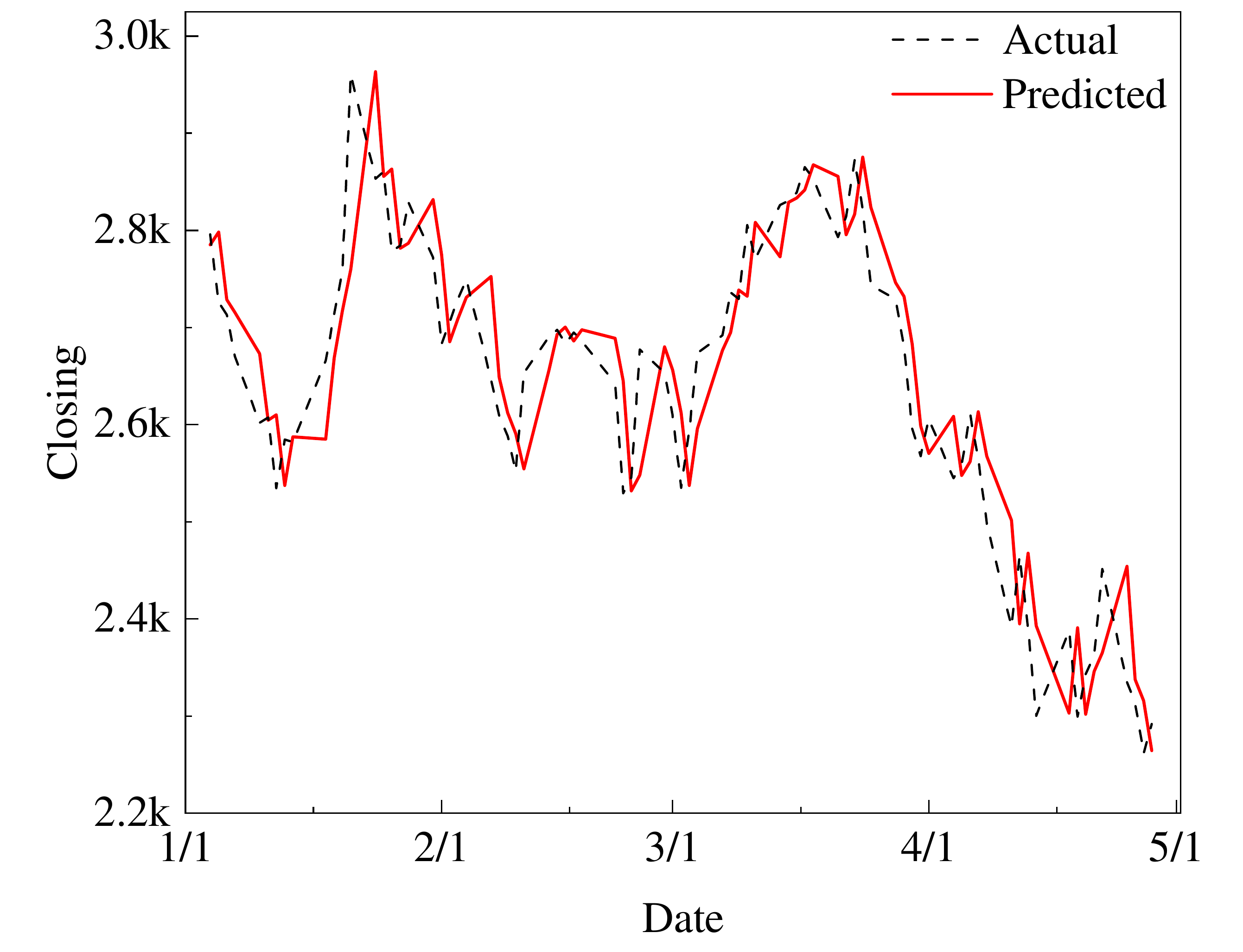} 
	}
	\quad
	\subfigure[]{
		\includegraphics[scale=0.15]{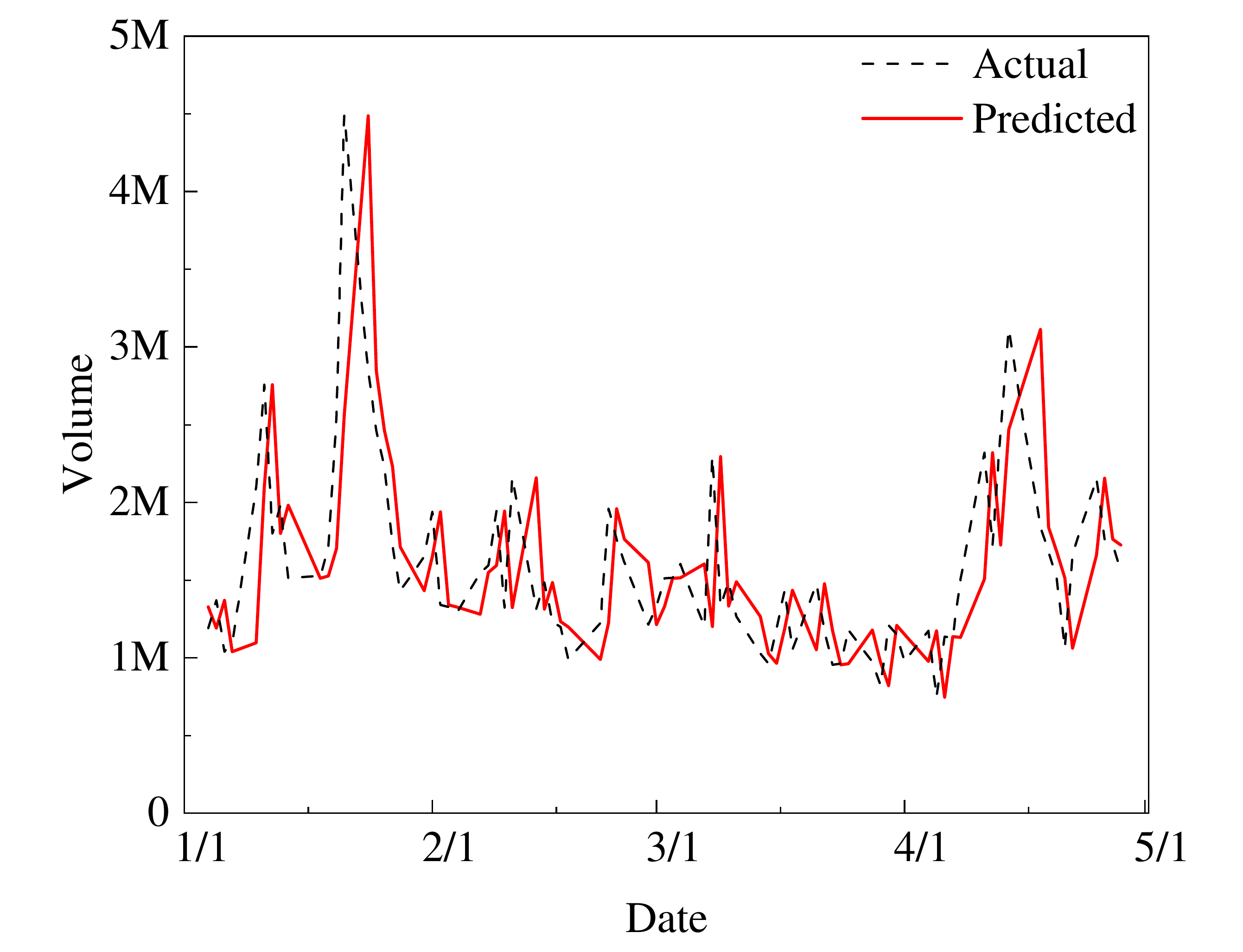} 
	}
	\caption{Curves of the stock price predicted by sQRNN: (a) the opening price, (b) the highest price, (c) the lowest price, (d) the closing price, and (e) the volume. }
\end{figure}
\end{document}